\documentclass[aps,prd,reprint,preprintnumbers,floatfix,superscriptaddress,nofootinbib,longbibliography]{revtex4-1}
\usepackage{graphicx}
\usepackage{xspace}
\usepackage{siunitx}
\usepackage{multirow}
\usepackage{comment}
\usepackage{amsmath}
\usepackage{braket}
\usepackage[normalem]{ulem}

\usepackage{lineno}
\makeatletter
\newcommand{\oset}[3][0ex]{%
  \mathrel{\mathop{#3}\limits^{
    \vbox to#1{\kern-2.5\ex@
    \hbox{$\scriptstyle#2$}\vss}}}}
\makeatother

\newcommand\overbar[1]{\oset[-0.2ex]{%
   \textbf{--}}{#1}}

\newcommand{\nue}{\ensuremath{\nu_{e}}\xspace}
\newcommand{\nuecc}{\ensuremath{\nu_{e}}~CC\xspace}
\newcommand{\nuebar}{\ensuremath{\overbar{\nu}_e}\xspace}
\newcommand{\nuebarcc}{\ensuremath{\overbar{\nu}_e}~\text{CC}\xspace}
\newcommand{\nueParen}{\ensuremath{{\nu}_{e}~(\bar{\nu}_{e})}\xspace}
\newcommand{\nueParenCC}{\ensuremath{{\nu_{e}~(\bar{\nu}_{e})}}~\text{CC}\xspace}

\newcommand{\numu}{\ensuremath{\nu_{\mu}}\xspace}
\newcommand{\numucc}{\ensuremath{\nu_{\mu}}~\text{CC}\xspace}
\newcommand{\numubar}{\ensuremath{\overbar{\nu}_\mu}\xspace}
\newcommand{\numubarcc}{\ensuremath{\overbar{\nu}_\mu}~\text{CC}\xspace}
\newcommand{\numuParen}{\ensuremath{{\nu}_{\mu}~(\bar{\nu}_{\mu})}\xspace}
\newcommand{\numuParenCC}{\ensuremath{{\nu_{\mu}~(\bar{\nu}_{\mu})}}~\text{CC}\xspace}

\newcommand{\nutau}{\ensuremath{\nu_{\tau}}\xspace}

\newcommand{\LL}{\ensuremath{-2\ln \cal{L}}\xspace}
\newcommand{\dmsq}{\ensuremath{\Delta m^2_{32}}\xspace}
\newcommand{\snsq}{\ensuremath{\sin^2 \theta_{23}}\xspace}
\newcommand{\deltacp}{\ensuremath{\delta_{\rm{CP}}}\xspace}
\newcommand{\qsq}{\ensuremath{Q^{2}}\xspace}

\newcommand{\pt}{\ensuremath{p_{\rm T}}\xspace}

\newcommand{\enu}{\ensuremath{E_{\nu}}\xspace}
\newcommand{\ehad}{\ensuremath{E_{\rm had}}\xspace}
\newcommand{\efrac}{\ensuremath{E_{\rm frac}}\xspace}

\newcommand{\evtcnn}{\ensuremath{\text{CNN}_{\text{evt}}}\xspace}

\newcommand{\valencia}{Val\`{e}ncia}

\usepackage[dvipsnames]{xcolor}
\usepackage{xspace, tabularx}
\usepackage[caption=false]{subfig}

\usepackage{threeparttable}

\begin{document}
\preprint{FERMILAB-PUB-21-373-ND}
\title{
Improved measurement of neutrino oscillation parameters by the NOvA experiment
}
\newcommand{\ANL}{Argonne National Laboratory, Argonne, Illinois 60439, 
USA}
\newcommand{\ICS}{Institute of Computer Science, The Czech 
Academy of Sciences, 
182 07 Prague, Czech Republic}
\newcommand{\IOP}{Institute of Physics, The Czech 
Academy of Sciences, 
182 21 Prague, Czech Republic}
\newcommand{\Atlantico}{Universidad del Atlantico,
Carrera 30 No. 8-49, Puerto Colombia, Atlantico, Colombia}
\newcommand{\BHU}{Department of Physics, Institute of Science, Banaras 
Hindu University, Varanasi, 221 005, India}
\newcommand{\UCLA}{Physics and Astronomy Department, UCLA, Box 951547, Los 
Angeles, California 90095-1547, USA}
\newcommand{\Caltech}{California Institute of 
Technology, Pasadena, California 91125, USA}
\newcommand{\Cochin}{Department of Physics, Cochin University
of Science and Technology, Kochi 682 022, India}
\newcommand{\Charles}
{Charles University, Faculty of Mathematics and Physics,
 Institute of Particle and Nuclear Physics, Prague, Czech Republic}
\newcommand{\Cincinnati}{Department of Physics, University of Cincinnati, 
Cincinnati, Ohio 45221, USA}
\newcommand{\CSU}{Department of Physics, Colorado 
State University, Fort Collins, CO 80523-1875, USA}
\newcommand{\CTU}{Czech Technical University in Prague,
Brehova 7, 115 19 Prague 1, Czech Republic}
\newcommand{\Dallas}{Physics Department, University of Texas at Dallas,
800 W. Campbell Rd. Richardson, Texas 75083-0688, USA}
\newcommand{\DallasU}{University of Dallas, 1845 E 
Northgate Drive, Irving, Texas 75062 USA}
\newcommand{\Delhi}{Department of Physics and Astrophysics, University of 
Delhi, Delhi 110007, India}
\newcommand{\JINR}{Joint Institute for Nuclear Research,  
Dubna, Moscow region 141980, Russia}
\newcommand{\Erciyes}{
Department of Physics, Erciyes University, Kayseri 38030, Turkey}
\newcommand{\FNAL}{Fermi National Accelerator Laboratory, Batavia, 
Illinois 60510, USA}
\newcommand{\UFG}{Instituto de F\'{i}sica, Universidade Federal de 
Goi\'{a}s, Goi\^{a}nia, Goi\'{a}s, 74690-900, Brazil}
\newcommand{\Guwahati}{Department of Physics, IIT Guwahati, Guwahati, 781 
039, India}
\newcommand{\Harvard}{Department of Physics, Harvard University, 
Cambridge, Massachusetts 02138, USA}
\newcommand{\Houston}{Department of Physics, 
University of Houston, Houston, Texas 77204, USA}
\newcommand{\IHyderabad}{Department of Physics, IIT Hyderabad, Hyderabad, 
502 205, India}
\newcommand{\Hyderabad}{School of Physics, University of Hyderabad, 
Hyderabad, 500 046, India}
\newcommand{\IIT}{Illinois Institute of Technology,
Chicago IL 60616, USA}
\newcommand{\Indiana}{Indiana University, Bloomington, Indiana 47405, 
USA}
\newcommand{\INR}{Institute for Nuclear Research of Russia, Academy of 
Sciences 7a, 60th October Anniversary prospect, Moscow 117312, Russia}
\newcommand{\Iowa}{Department of Physics and Astronomy, Iowa State 
University, Ames, Iowa 50011, USA}
\newcommand{\Irvine}{Department of Physics and Astronomy, 
University of California at Irvine, Irvine, California 92697, USA}
\newcommand{\Jammu}{Department of Physics and Electronics, University of 
Jammu, Jammu Tawi, 180 006, Jammu and Kashmir, India}
\newcommand{\Lebedev}{Nuclear Physics and Astrophysics Division, Lebedev 
Physical 
Institute, Leninsky Prospect 53, 119991 Moscow, Russia}
\newcommand{\Magdalena}{Universidad del Magdalena, Carrera 32 No 22-08 Santa Marta, Colombia}
\newcommand{\MSU}{Department of Physics and Astronomy, Michigan State 
University, East Lansing, Michigan 48824, USA}
\newcommand{\Crookston}{Math, Science and Technology Department, University 
of Minnesota Crookston, Crookston, Minnesota 56716, USA}
\newcommand{\Duluth}{Department of Physics and Astronomy, 
University of Minnesota Duluth, Duluth, Minnesota 55812, USA}
\newcommand{\Minnesota}{School of Physics and Astronomy, University of 
Minnesota Twin Cities, Minneapolis, Minnesota 55455, USA}
\newcommand{\Mississippi}{University of Mississippi, University, Mississippi 38677, USA}
\newcommand{\NISER}{National Institute of Science Education and Research,
Khurda, 752050, Odisha, India}
\newcommand{\Oxford}{Subdepartment of Particle Physics, 
University of Oxford, Oxford OX1 3RH, United Kingdom}
\newcommand{\Panjab}{Department of Physics, Panjab University, 
Chandigarh, 160 014, India}
\newcommand{\Pitt}{Department of Physics, 
University of Pittsburgh, Pittsburgh, Pennsylvania 15260, USA}
\newcommand{\QMU}{School of Physics and Astronomy,
 Queen Mary University of London,
London E1 4NS, United Kingdom}
\newcommand{\RAL}{Rutherford Appleton Laboratory, Science 
and 
Technology Facilities Council, Didcot, OX11 0QX, United Kingdom}
\newcommand{\SAlabama}{Department of Physics, University of 
South Alabama, Mobile, Alabama 36688, USA} 
\newcommand{\Carolina}{Department of Physics and Astronomy, University of 
South Carolina, Columbia, South Carolina 29208, USA}
\newcommand{\SDakota}{South Dakota School of Mines and Technology, Rapid 
City, South Dakota 57701, USA}
\newcommand{\SMU}{Department of Physics, Southern Methodist University, 
Dallas, Texas 75275, USA}
\newcommand{\Stanford}{Department of Physics, Stanford University, 
Stanford, California 94305, USA}
\newcommand{\Sussex}{Department of Physics and Astronomy, University of 
Sussex, Falmer, Brighton BN1 9QH, United Kingdom}
\newcommand{\Syracuse}{Department of Physics, Syracuse University,
Syracuse NY 13210, USA}
\newcommand{\Tennessee}{Department of Physics and Astronomy, 
University of Tennessee, Knoxville, Tennessee 37996, USA}
\newcommand{\Texas}{Department of Physics, University of Texas at Austin, 
Austin, Texas 78712, USA}
\newcommand{\Tufts}{Department of Physics and Astronomy, Tufts University, Medford, 
Massachusetts 02155, USA}
\newcommand{\UCL}{Physics and Astronomy Department, University College 
London, 
Gower Street, London WC1E 6BT, United Kingdom}
\newcommand{\Virginia}{Department of Physics, University of Virginia, 
Charlottesville, Virginia 22904, USA}
\newcommand{\WSU}{Department of Mathematics, Statistics, and Physics,
 Wichita State University, 
Wichita, Kansas 67206, USA}
\newcommand{\WandM}{Department of Physics, William \& Mary, 
Williamsburg, Virginia 23187, USA}
\newcommand{\Wisconsin}{Department of Physics, University of 
Wisconsin-Madison, Madison, Wisconsin 53706, USA}
\newcommand{\deceased}{Deceased.}
\affiliation{\ANL}
\affiliation{\Atlantico}
\affiliation{\BHU}
\affiliation{\Caltech}
\affiliation{\Charles}
\affiliation{\Cincinnati}
\affiliation{\Cochin}
\affiliation{\CSU}
\affiliation{\CTU}
\affiliation{\Delhi}
\affiliation{\Erciyes}
\affiliation{\FNAL}
\affiliation{\UFG}
\affiliation{\Guwahati}
\affiliation{\Harvard}
\affiliation{\Houston}
\affiliation{\Hyderabad}
\affiliation{\IHyderabad}
\affiliation{\IIT}
\affiliation{\Indiana}
\affiliation{\ICS}
\affiliation{\INR}
\affiliation{\IOP}
\affiliation{\Iowa}
\affiliation{\Irvine}
\affiliation{\JINR}
\affiliation{\Lebedev}
\affiliation{\Magdalena}
\affiliation{\MSU}
\affiliation{\Duluth}
\affiliation{\Minnesota}
\affiliation{\Mississippi}
\affiliation{\NISER}
\affiliation{\Panjab}
\affiliation{\Pitt}
\affiliation{\QMU}
\affiliation{\SAlabama}
\affiliation{\Carolina}
\affiliation{\SMU}
\affiliation{\Stanford}
\affiliation{\Sussex}
\affiliation{\Syracuse}
\affiliation{\Texas}
\affiliation{\Tufts}
\affiliation{\UCL}
\affiliation{\Virginia}
\affiliation{\WSU}
\affiliation{\WandM}
\affiliation{\Wisconsin}

\author{M.~A.~Acero}
\affiliation{\Atlantico}

\author{P.~Adamson}
\affiliation{\FNAL}



\author{L.~Aliaga}
\affiliation{\FNAL}






\author{N.~Anfimov}
\affiliation{\JINR}


\author{A.~Antoshkin}
\affiliation{\JINR}


\author{E.~Arrieta-Diaz}
\affiliation{\Magdalena}

\author{L.~Asquith}
\affiliation{\Sussex}


\author{A.~Aurisano}
\affiliation{\Cincinnati}


\author{A.~Back}
\affiliation{\Indiana}
\affiliation{\Iowa}

\author{C.~Backhouse}
\affiliation{\UCL}

% ...

\collaboration{The NOvA Collaboration}
\noaffiliation


\begin{abstract}
We present new 
\numu{}\,$\rightarrow$\,\nue, \numu{}\,$\rightarrow$\,\numu,  \numubar{}\,$\rightarrow$\,\nuebar, and \numubar{}\,$\rightarrow$\,\numubar oscillation measurements by the NOvA experiment, with a 50\% increase in neutrino-mode beam exposure over the previously reported results.
The additional data,  combined with previously published neutrino and antineutrino data,
are all analyzed using improved techniques and simulations.
A joint fit to the \nue, \numu, \nuebar, and \numubar  candidate samples within the 3-flavor neutrino oscillation framework continues to yield a best-fit point in the normal mass ordering and the upper octant of the $\theta_{23}$ mixing angle, with $\dmsq = (2.41\pm0.07)\times 10^{-3}$~eV$^2$ and $\snsq = 0.57^{+0.03}_{-0.04}$. 
The data disfavor combinations of oscillation parameters that give rise to a large asymmetry in the rates of \nue and \nuebar appearance. This includes  values of the CP-violating phase in the vicinity of $\deltacp = \pi/2$ which are excluded by $> 3\,\sigma$ for the inverted mass ordering, and values around $\deltacp = 3\pi/2$ in the normal ordering which are disfavored at 2\,$\sigma$ confidence. 

\end{abstract}





\maketitle


\section{\label{sec:intro} Introduction}

We report new measurements of neutrino oscillation parameters using neutrino and antineutrino data from the NOvA experiment. The data include a 50\% increase in neutrino-mode beam exposure over the previously reported results~\cite{Acero:2019ksn}. We perform a joint fit to \numuParen{}\,$\rightarrow$\,\nueParen and \numuParen{}\,$\rightarrow$\,\numuParen oscillations utilizing improvements in the analysis of these data.

Numerous experiments~\cite{Fukuda:1998mi,Fukuda:2002pe,Ahmad:2002jz,Eguchi:2002dm,Michael:2006rx,Abe:2011sj,Abe:2011fz,An:2012eh,Ahn:2012nd} corroborate the paradigm in which three neutrino mass eigenstates ($\nu_1, \nu_2, \nu_3$) mix to form the three flavor eigenstates (\nue, \numu, \nutau). The mixing can be expressed by the unitary matrix, $U_\text{PMNS}$, named for Pontecorvo, Maki, Nakagawa, and Sakata. $U_\text{PMNS}$ can be parameterized by three mixing angles ($\theta_{12}, \theta_{23}, \theta_{13}$) along with a phase (\deltacp) that, if different from 0 or $\pi$, indicates violation of Charge-Parity (CP) symmetry. Neutrino mixing gives rise to oscillations from one flavor state to another, dependent on the mixing parameters and the mass splittings ($\Delta m^2_{ij} \equiv m_{i}^2 - m_{j}^2$). 

Using the definition of $\nu_1$ as having the largest \nue contribution, it has been established that $\Delta m^2_{21}$ is positive, and therefore, the $\nu_2$ mass eigenstate is heavier than $\nu_1$. However, the sign of the larger mass splitting, $\Delta m^2_{32}$, is unknown. If this term is positive, then the third mass eigenstate is the heaviest, and the mass ordering is labeled as the Normal Ordering (NO) (also referred to as Normal Hierarchy). The alternative is referred to as Inverted Ordering (IO) (or Inverted Hierarchy). 
Knowing the mass ordering would constrain models of neutrino masses~\cite{Mohapatra:2006gs,Nunokawa:2007qh,Altarelli:2010gt,King:2015aea,Petcov:2017ggy} and could aid in the resolution of the Dirac or Majorana nature of the neutrino~\cite{Pascoli:2002xq,Bahcall:2004ip}.

The mass ordering affects the rates of \numu{}\,$\rightarrow$\,\nue and \numubar{}\,$\rightarrow$\,\nuebar oscillations when neutrinos travel through the Earth as compared to a vacuum. 
Coherent forward scattering on electrons in the Earth's crust enhances the rate of \numu{}\,$\rightarrow$\,\nue oscillations and suppresses \numubar{}\,$\rightarrow$\,\nuebar for the NO while the enhancement and suppression is reversed for the IO. This matter effect~\cite{Wolfenstein:1977ue} changes the oscillation probabilities for NOvA by $\sim20\%$. Depending on the value of \deltacp and the mass ordering itself, NOvA may be able to exploit the resulting neutrino-antineutrino asymmetry to measure the sign of $\Delta m^2_{32}$ and thus determine the mass ordering. 

NOvA also has sensitivity to \deltacp, which will increase the \numu{}\,$\rightarrow$\,\nue oscillation probability if $\text{sin}\,\deltacp$ is positive and suppress oscillations if negative (the effect is reversed for antineutrinos). Additionally, a non-zero value of $\sin\deltacp$ would identify the neutrino sector as a source of CP violation which is central to some explanations of the matter-antimatter asymmetry observed based on leptogenesis ~\cite{Fukugita:1986hr,Buchmuller:1996pa,Buchmuller:2004nz,Buchmuller:2005eh,Pilaftsis:1997jf}. Since a measurement of both the mass ordering and \deltacp rely on a comparison of \nue and \nuebar appearance, certain combinations of \deltacp and mass ordering will be degenerate with others for NOvA's oscillation baseline. 

Finally, the angle $\theta_{23}$ largely determines the coupling of the \numu and \nutau states to the $\nu_3$ mass state.  In the case of maximal mixing, $\theta_{23} = \pi/4$, \numu and \nutau couple equally to $\nu_3$~\cite{Harrison:2002et}, which suggests a $\mu - \tau$ symmetry.  If non-maximal, $\theta_{23}$ could lie in the upper octant (UO, $\theta_{23} > \pi/4$) or lower octant (LO, $\theta_{23} < \pi/4$) with a stronger \numu or \nutau coupling, respectively.  Current measurements of $\theta_{23}$ are near maximal mixing~\cite{Acero:2019ksn,Michael:2006rx,Abe:2011sj}, but significant uncertainties remain making it the least precisely measured mixing angle.

Here, we reanalyze the data taken in the antineutrino-mode beam from June 29, 2016, to February 26, 2019, with an exposure of $12.5 \times 10^{20}$ protons on target (POT) delivered during \SI{321.1}{\second} of integrated beam-pulse time. These data are combined with an increased, and reanalyzed, neutrino-mode beam exposure of $13.6 \times 10^{20}$~POT from \SI{555.3}{\second} of integrated beam-pulse time recorded between February 6, 2014, to March 20, 2020. During these periods, the proton source achieved an average power of \SI{650}{\kilo\watt}, and a peak hourly-averaged power of \SI{756}{\kilo\watt}. 

In addition to the increased neutrino-mode beam exposure, this analysis introduces various improvements that will be described in detail in the following sections. There are changes to the underlying neutrino interaction simulation, particle propagation, and detector response models. The reconstruction uses a new clustering algorithm and expands the use of neural networks. Furthermore, the Near-to-Far extrapolation method has been expanded to further constrain the FD predictions, which also reduces the impact of systematic uncertainties on the analysis by up to 9\% as compared to the previous method. Finally, we have improved some systematic uncertainties and introduced new ones associated with the above changes.

\section{\label{sec:nova-sim-calib} The NO\lowercase{v}A Experiment and Simulations}

NOvA observes \numuParen{}\,$\rightarrow$\,\nueParen appearance and \numuParen{}\,$\rightarrow$\,\numuParen disappearance oscillations using two functionally-identical tracking calorimeters~\cite{Ayres:2007tu} deployed in Fermilab's NuMI beam~\cite{Adamson:2015dkw}.
Charged particle tracking is accomplished via PVC cells filled with a mineral oil-based liquid scintillator~\cite{Mufson:2015kga}. The cells are \SI{6.6}{\centi\meter}\,$\times$\,\SI{3.9}{\centi\meter} in cross section and are oriented in alternating vertical and horizontal planes to achieve 3D reconstruction. The \SI{290}{ton} Near Detector (ND) is located \SI{100}{\meter} underground and $\sim$\SI{1}{\kilo\meter} from the production target. The main body of the ND is followed by a muon range stack where the active planes are interleaved with steel plates. The \SI{14}{\kilo ton} Far Detector (FD) is located at Ash River, Minnesota, $\sim$\SI{810}{\kilo\meter} from the source. Being located on the surface with a modest rock overburden, the FD receives a cosmic-ray flux of \SI{130}{\kilo\hertz}. This analysis benefits from an updated simulation of the geometries of the detectors and their surroundings that more accurately reflects the surrounding rock composition and detectors as built. 

Both detectors are centered \SI{14.6}{\milli\radian} off the beam axis and receive a narrow-band neutrino flux peaked at \SI{1.8}{\giga\eV}. Magnetic focusing horns are used to select the sign of the neutrino parents, producing a 93\% (92\%) pure \numu (\numubar) beam between \SIrange{1}{5}{\giga\eV}. The majority of contamination is due to ``wrong-sign'' neutrinos (\textit{i.e.} \numubar in a \numu selected beam and vice versa). The neutrino flux delivered to the detectors is calculated using \textsc{geant4}-based simulations of particle production and transport through the beamline components~\cite{Adamson:2015dkw,Agostinelli:2002hh} reweighted to incorporate external measurements using the Package to Predict the Flux (PPFX)~\cite{Aliaga:2016oaz,Paley:2014rpb,Alt:2006fr,Abgrall:2011ae,Barton:1982dg,Seun:2007zz,Tinti:2010zz,Lebedev:2007zz,Baatar:2012fua,Skubic:1978fi,Denisov:1973zv,Carroll:1978hc,Abe:2012av,Gaisser:1975et,Cronin:1957zz,Allaby:1969de,Longo:1962zz,Bobchenko:1979hp,Fedorov:1977an,Abrams:1969jm}.

Neutrino interactions are simulated using a custom model configuration of \textsc{genie} 3.0.6 \cite{Andreopoulos:2009rq,Andreopoulos:2015wxa} tuned to external and NOvA ND data.\footnote{Neutrino interactions in this analysis were inadvertently simulated with event kinematics of GENIE configuration N18\_10j\_00\_000 but integrated rates with configuration N18\_10j\_02\_11a.  These two configurations have the same model set and differ only in the tune of the resonant, non-resonant background, and DIS free nucleon cross sections, where the N18\_10j\_00\_000 tune used inclusive neutrino scattering data and the N18\_10j\_02\_11a tune used $1\pi$ and $2\pi$  production in addition to the inclusive neutrino scattering data ~\cite{GENIE:2021zuu,Tena-Vidal:2018misc}. The predicted Far Detector event spectra generated using N18\_10j\_02\_11a are consistent with the predictions used in this measurement within the systematic uncertainties.} In this configuration, charged-current (CC) quasi-elastic (QE) scattering is simulated using the model of Nieves \textit{et al.}~\cite{Nieves:2004wx}, which includes the effects of long-range nucleon correlations calculated according to the Random Phase Approximation (RPA)~\cite{Martini:2009uj,Nieves:2004wx,Pandey:2014tza}.  The CCQE axial vector form factor is a $z$-expansion parameterization tuned to neutrino-deuterium scattering data \cite{Meyer:2016oeg}.  CC interactions with two nucleons producing two holes (2p2h) are given by the IFIC \valencia{} model~\cite{Nieves:2011pp,Gran:2013kda}.  The initial nuclear state is represented by a local Fermi gas in both the QE and 2p2h models, and by a global relativistic Fermi gas for all other processes.
Baryon resonance (RES) and coherent pion production are simulated using the Berger-Sehgal models with final-state mass effects taken into account \cite{Berger:2007rq,Berger:2008xs}. 
Deep inelastic scattering (DIS) and non-resonant background below the DIS region are described using the Bodek-Yang model \cite{Bodek:2002ps} with hadronization simulated by a data-driven parameterization \cite{Yang:2009zx} coupled to \textsc{pythia} \cite{Sjostrand:2006za}.  Bare nucleon cross-sections for RES, DIS, and non-resonant background processes are tuned by \textsc{genie} to neutrino scattering data.  Final state interactions (FSI) are simulated by the \textsc{genie} hN semi-classical intranuclear cascade model in which pion interaction probabilities are assigned according to Oset \textit{et al.}~\cite{Salcedo:1987md} and pion-nucleon scattering data.


The 2p2h and FSI models in this \textsc{genie} configuration are adjusted to produce a NOvA-specific neutrino interaction model tune.  The 2p2h model is fit to \numu CC inclusive scattering data from the NOvA ND. Inspired by Gran \textit{et al.}~\cite{Gran:2018fxa}, this 2p2h tune enhances the base model as a function of energy and momentum transfer to the nucleus and is applied to all CC 2p2h interactions for both the neutrino and anti-neutrino beams.  The parameters governing $\pi^{\pm}$ and $\pi^{0}$ FSI are adjusted to obtain agreement with $\pi^{+}$ on ${^{12}\text{C}}$ scattering data~\cite{Allardyce:1973ce,Saunders:1996ic,Meirav:1988pn,Levenson:1983xu,Ashery:1981tq,Ashery:1984ne,PinzonGuerra:2016uae}.

The propagation of final state particles through the detectors is simulated by an updated version of \textsc{geant4} (v10.4) \cite{Geant:2017ats}, which provides the input for the detector response simulation~\cite{Aurisano:2015oxj}.  In addition, a custom patch to the new version implements an exact calculation of the density effect correction to the Bethe equation using Sternheimer's method~\cite{Sternheimer:1952jn} as opposed to the approximate parameterization used previously (a 1\% or less change to the muon range and energy lost in dead material). 

The absolute energy scale for both detectors is calibrated using the minimum ionizing portion of stopping cosmic-ray muon tracks \cite{Singh:2019lqg}.  
The calibration procedure is now applied separately to the data in shorter time periods to account for an observed 0.3\% decrease in detected light per year.

\section{\label{sec:event-reco-select-energy} Reconstruction and Selection}

\footnotetext{The FD sample efficiency, purity, and energy resolution are based on the simulated event samples at the determined best-fit point. Energy resolution is defined as the RMS of the distribution: $1~-~E_{\nu}^{reco}/E_{\nu}^{true}$.  Wrong-sign events are treated as background for the \nueParenCC samples and signal for the \numuParenCC samples. For the efficiency calculations, the denominator is the number of true signal interactions in the detector with no other selection criteria applied.}

The first stage of reconstruction is to group hits, which are measurements of deposited energy in a cell above a preset threshold, into single-neutrino-interaction events. This clustering, performed based on hit proximity in time and space, now uses a new method that reduces the rate of mis-clustered hits in the high occupancy environment of the ND \cite{Pershey:2018gtf}. Mis-clustering had previously led to differences in data-MC selection efficiency, which are now reduced to the sub-percent level. The other reconstruction techniques remain unchanged from the previous analysis~\cite{Acero:2019ksn}.

For each event, initial selections are applied to ensure basic data quality.
Additionally, events are required to be sufficiently far from the edges of the detector such that energy is not lost to exiting final-state particles, and so entering background events are not selected as signal. These containment criteria have been re-optimized for this analysis due to changes in the geometry model and hit grouping algorithm, but follow the same outline as described in Ref.~\cite{Acero:2019ksn}. 

A convolutional neural network, \evtcnn~\cite{Aurisano:2016jvx}, is used to classify neutrino event candidates into \nue CC, \numu CC, NC, or cosmogenic background. The network is trained using simulated calibrated hits that have been clustered into single neutrino interactions, as well as cosmogenic data. Scores from \evtcnn are used to create two non-overlapping samples of either inclusive \numuParenCC or \nueParenCC candidate events.  Updates to this algorithm provide improved performance and decreased dependency on calorimetric energy, the dominant source of systematic uncertainty in the results presented here. This is achieved by scaling up or down the energy of all hits while training the CNN. The scale factors used are drawn on an event-by-event basis from a normal distribution with a 1 $\sigma$ range from 0.9\,-\,1.1 \cite{Groh:2021qyu}. This training procedure reduced the influence of calibration uncertainties on \evtcnn classification decisions to a negligible level.

\begin{table}{}
\caption[]{
FD energy resolution and purity$^2$, in the selected energy ranges (\SIrange{0}{5}{\giga\eV} for \numu and \SIrange{0}{4}{\giga\eV} for \nue), for the subsamples used in the Near-to-Far extrapolation and oscillation fits. \efrac for the \numu samples is defined in the text. The \nueParen peripheral is a rate-only sample, therefore,  \enu is not determined.}
\label{tb:sample-res-pur}
\begin{tabular}{c|ccc}
	\hline
	\hline
    \multicolumn{2}{c}{Sample bins} & Energy res. & Sample purity \\
	\hline
	\multirow{4}{*}{\nueParen}  
	\rule[-4pt]{0pt}{13pt}
    & Core, Low \evtcnn  & 14.1\% (13.7\%) & 51\% (36\%) \\
    & Core, High \evtcnn & 9.4\% (8.9\%)   & 79\% (69\%) \\
    & Peripheral         & --              & 57\%~(43\%) \\
    \cline{2-4}
    \rule[-4pt]{0pt}{13pt}
    & Combined              & 10.7\% (8.8\%)  & 69\% (58\%) \\
    \hline
    \multirow{5}{*}{\numuParen}
    \rule[-4pt]{0pt}{13pt}
    & 1 (lowest \efrac)  & 7.8\% (8.5\%)   & 99\% (99\%)   \\
    & 2                  & 9.2\% (8.9\%)   & 99\% (99\%)   \\
    & 3                  & 10.4\% (9.7\%)  & 97\% (98\%)   \\
    & 4 (highest \efrac) & 11.5\% (10.2\%) & 92\% (95\%)   \\
    \cline{2-4}
    \rule[-4pt]{0pt}{13pt}
    & Combined              & 9.1\% (8.2\%)   & 96\% (98\%)   \\
    \hline
    \hline
    \end{tabular}
\end{table}


Effective rejection of cosmogenic backgrounds at the FD is paramount due to the significant flux of cosmic-ray particles it receives. A new CNN, trained to identify cosmogenic backgrounds has been introduced, is applied in parallel to cosmic-identifying boosted decision trees (BDTs). The BDTs have been trained on samples selected to contain signal-like cosmogenic particles. Together the CNN and BDTs reduce the cosmic contamination in the selected samples to $<$\,5\%, a total reduction of 6 orders of magnitude, comparable to the previous analysis. For fully contained \nue events, the BDT replaces the previous cosmic rejection method, which directly used reconstructed position and kinematic event information. 

Neutrino energy, \enu, is determined using different methods for the \nue and \numu CC candidate events.
The energies of \nue CC candidates are parameterized using a quadratic function determined from a 2D fit to the simulated electromagnetic (EM) and hadronic calorimetric energies ($E_{EM}$ and $E_{had}$ respectively). The two components produce different detector responses and are separated using a third CNN classifier that identifies EM-like hit clusters within the event with the remaining clusters being classified as hadronic~\cite{Psihas:2018czu}. 
For \numu CC candidates, \enu is the sum of the muon energy, determined by the track length, and the total calorimetric energy of the hadronic system, \ehad. The muon is identified with a BDT that utilizes track length, multiple Coulomb scattering, and energy deposition, while the hadronic system is taken as all hits not associated with the muon track.

The selection criteria and energy estimation techniques were developed based on ND beam and FD cosmic data, along with simulated samples prior to inspecting the FD beam data distributions. The algorithms were trained separately on neutrino and antineutrino beam modes due to differences in beam purity and interactions. 

The sensitivity of the oscillation fit is enhanced by splitting the fully contained \nue and \nuebar CC, ``core'',  samples into low and high purity bins, based on the scores output by \evtcnn.
At the FD, the \nueParen selection efficiency for signal events in the core sample is 54\% (64\%)\footnotemark[\value{footnote}]. To further increase the efficiency of the FD sample, a ``peripheral'' selection is included, consisting of events that fail the containment or cosmic rejection requirements but pass more strict selection criteria on the cosmic BDT and \evtcnn. This sample increases the total \nueParen selection efficiency to 63\% (75\%)$^2$ but is included only as an integrated rate in the oscillation fits due to possible energy bias caused by particles leaving the detector. Properties of these subsamples are summarized in Table~\ref{tb:sample-res-pur}.

For \numu CC candidates, the position and amplitude of the oscillation maximum in the FD energy spectra are strongly dependent on \dmsq and $\theta_{23}$, respectively. To maximize the sensitivity to these parameters, the candidates are divided into four equally populated samples based on the hadronic energy fraction, $\efrac=\ehad/\enu$, which is correlated with energy resolution and background contamination as summarized in Table~\ref{tb:sample-res-pur}. Sensitivity is further increased by using variably-sized \enu bins for these samples.

\section{\label{sec:near-to-far} Near-to-Far Extrapolation}

\begin{figure}[t]
    \makebox[\columnwidth][c]{
        \subfloat[]{\label{fig:nd-numu-data-figure}\includegraphics[width=1.751in]{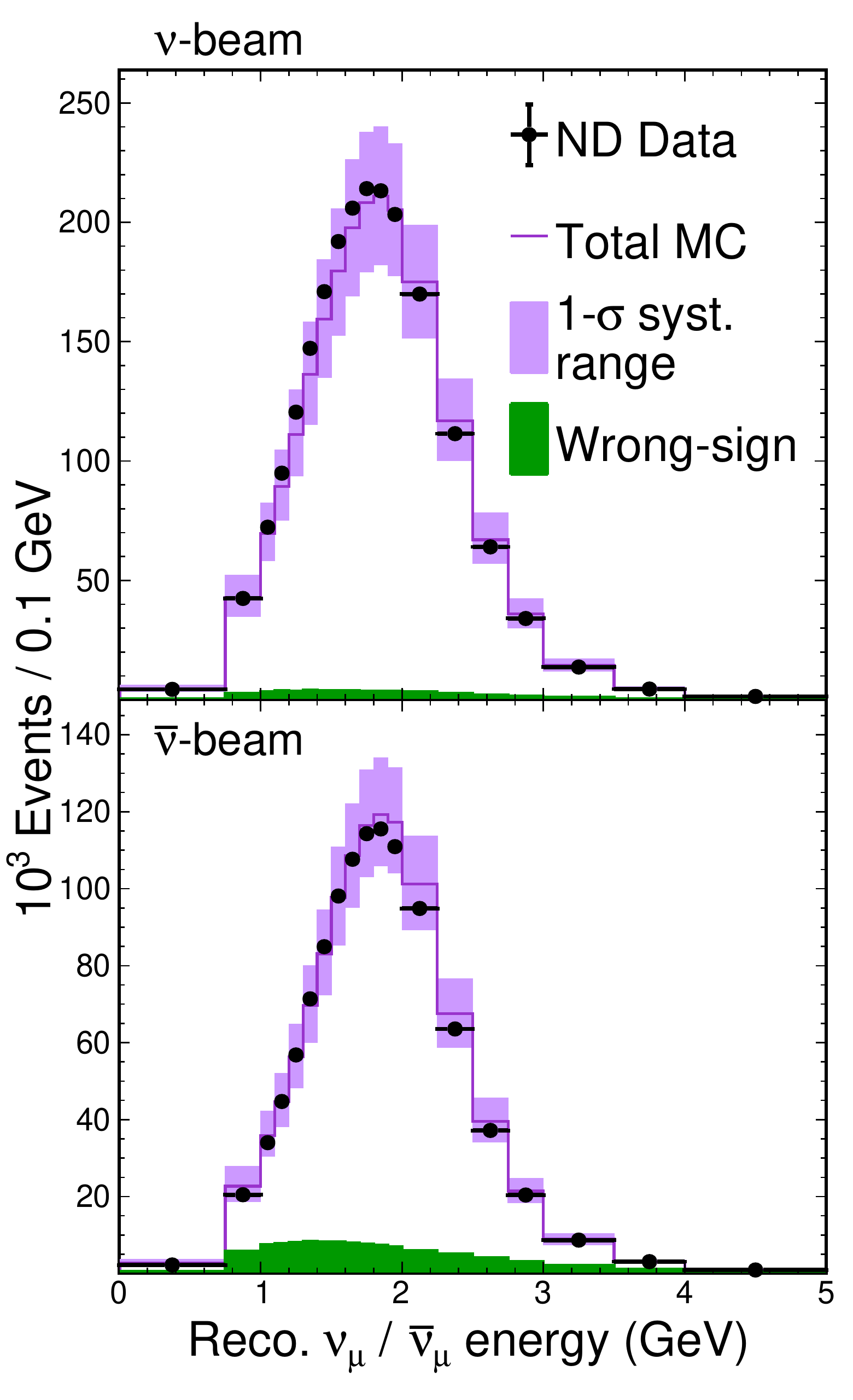}}
        \subfloat[]{\label{fig:nd-nue-data-figure}\includegraphics[width=1.751in]{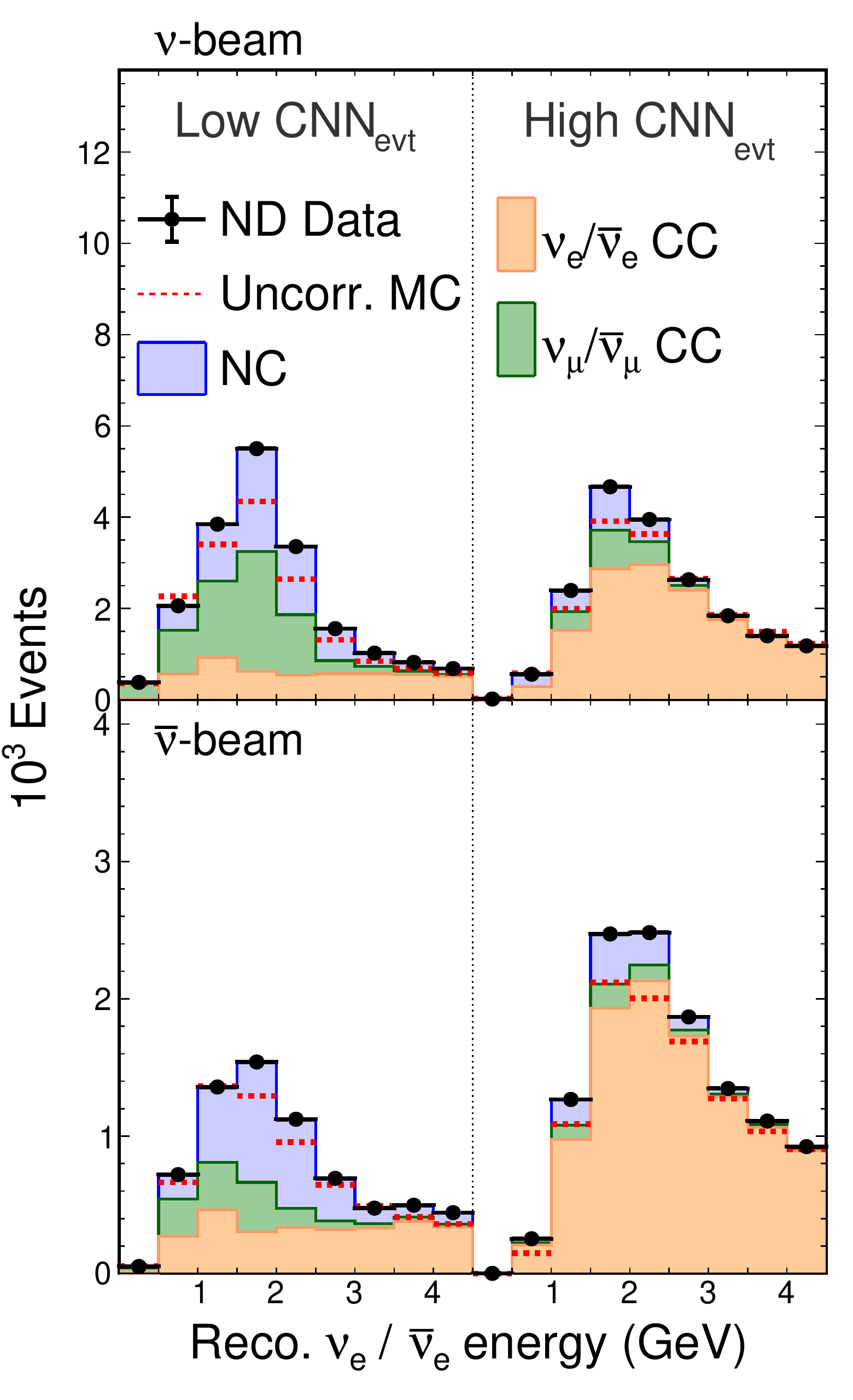}}
    }
    \caption{
        Reconstructed neutrino energy spectra for the (a) ND \numucc and (b) ND \nuecc samples with the neutrino-mode beam on top and antineutrino-mode on the bottom~\cite{supplemental-numu}. The \numucc \efrac sub-samples have been combined. The \nuecc low and high \evtcnn sub-samples are shown. 
        Dashed lines in the ND \nue and \nuebar spectra show the simulated counts before data-driven corrections and the colored regions show the breakdown by interaction type. 
    }
    \label{fig:nd-data-figure}
\end{figure}

This analysis extracts oscillation parameters using data-driven predictions of the FD spectra largely derived from high statistics measurements in the ND.
The \numuParen disappearance and \nueParen appearance signal spectra in the FD are predicted using the spectra of \numuParenCC candidate events in the ND (Fig.~\ref{fig:nd-numu-data-figure}).  The procedure begins with reweighting the simulation to obtain agreement with the data in each reconstructed \enu bin of the ND \numuParenCC candidate samples.  
Predicted rates of NC, \nue CC, and \nuebar CC interactions in the samples ($<$0.5\% total) are taken directly from the simulation and subtracted.  
The wrong-sign component of the samples (2.9\% and 10.5\% in the neutrino and antineutrino beams respectively) is also taken directly from the simulation.
The resulting corrected \numu$+$\xspace\numubar CC reconstructed \enu spectra are transformed to true \enu using the simulation. The spectra are then multiplied by the appropriate far-to-near ratios of the simulated samples in bins of true \enu.  This step accounts for beam divergence, differences in selection efficiency and acceptance between the two detectors, and the differences in the \numu and \nue cross sections.  Oscillation probabilities are applied to yield the predicted disappearance or appearance signal spectra in true \enu at the FD.  Matter effects are included in the oscillation probability calculations, with the Earth’s crust density assumed to be uniformly \SI[per-mode=symbol]{2.84}{\gram\per\centi\meter\cubed}~\cite{Bassin:2000ats}.  Finally, the predicted spectra are converted back to reconstructed \enu.

\nolinenumbers

\begin{figure}[t]
    \includegraphics[width=\columnwidth]{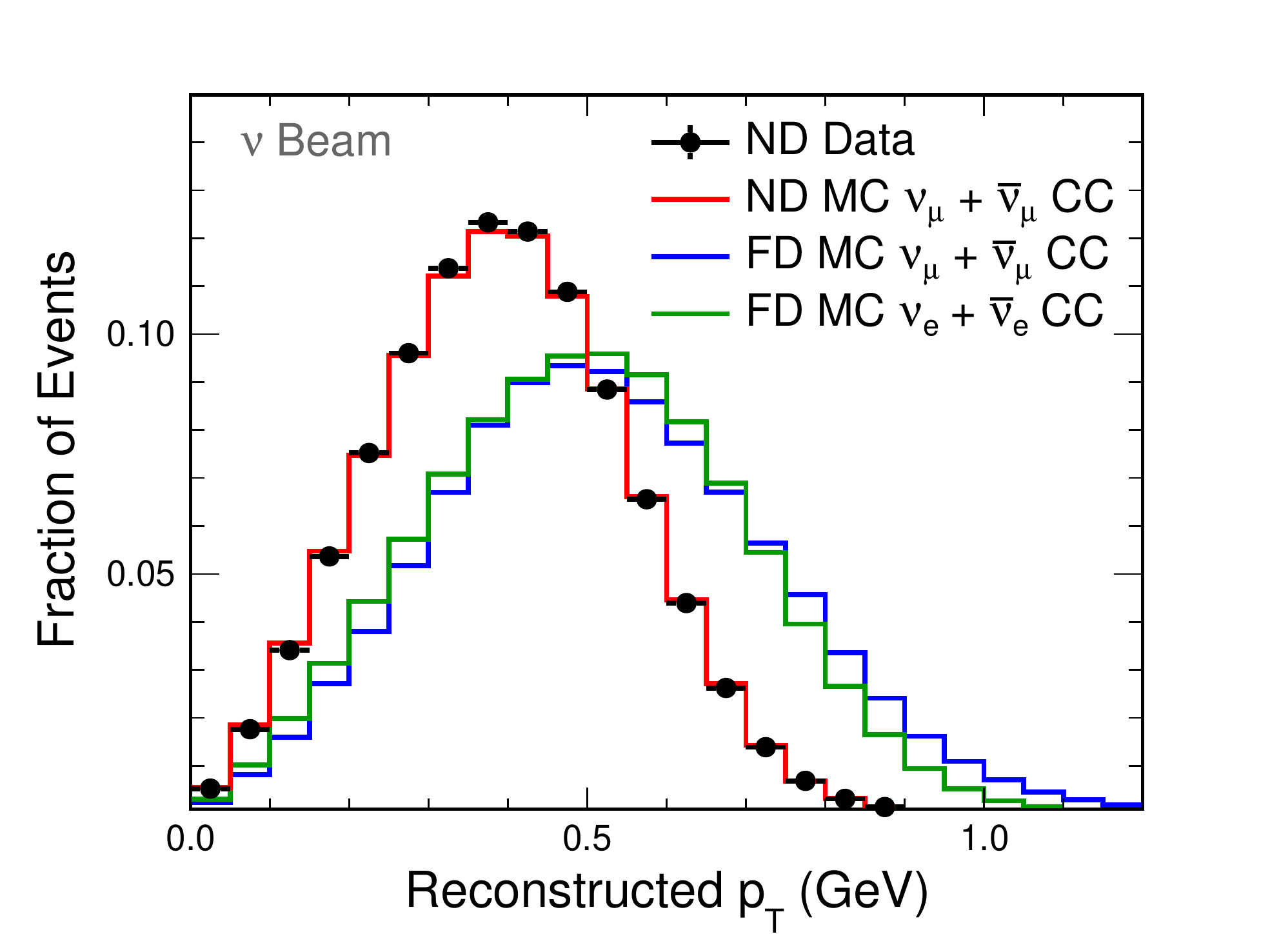}
\caption{
Distributions from the neutrino-mode beam of the fraction of selected events versus reconstructed \pt of the final state lepton, for the ND \numu CC data and simulation, and for the simulated FD \numu and \nue signal events.
The corresponding distributions from the antineutrino-mode beam are similar.}
\label{fig:pt_spectra}
\end{figure}

To reduce potential bias and the impact of uncertainties from the neutrino interaction model, the extrapolation to predict the disappearance and appearance signals is performed using variables in addition to \enu. As in the previous analysis, the extrapolations for the disappearance samples are done separately in each reconstructed hadronic energy fraction range (as given in Table~\ref{tb:sample-res-pur}), enabling neutrino interaction processes that occur in different inelasticity regions to be constrained independently.  In this analysis, the extrapolations for both disappearance and appearance samples are additionally performed separately in bins of reconstructed transverse momentum, \pt, of the final state charged lepton.  The smaller transverse extent of the ND leads to lower acceptance at higher \pt in the ND than in the FD (Fig.~\ref{fig:pt_spectra}), which results in the extrapolated predictions being sensitive to the modeling of the \pt-dependence of the neutrino interactions.  Extrapolating in bins of \pt reduces this sensitivity by enabling the ND data to constrain the \pt-dependence.  In the ND samples, the \pt bins divide each \enu bin into three equal populations for the extrapolation, and the resulting FD predictions are summed over the \pt bins for the oscillation fit.

Background spectra at the FD are also predicted using data-driven techniques. Cosmogenic backgrounds in both the appearance and disappearance samples are estimated using FD data collected outside the NuMI beam time window. Beam-induced backgrounds in the appearance samples are primarily CC interactions from the irreducible \nue$+$\xspace\nuebar component of the beam, with contributions from mis-identified NC and \numu$+$\xspace\numubar CC interactions.  The FD spectra for these backgrounds are predicted using the spectra of \nueParenCC candidate events in the ND (Fig.~\ref{fig:nd-nue-data-figure}).  Since the relative event rate between the ND and FD is different for the background components, the relative contribution of the different background components in the data needs to be estimated. In neutrino beam-mode these estimates are data-driven~\cite{NOvA:2018gge, Pershey:2018gtf} while they are taken directly form the simulation in antineutrino beam-mode.

\begin{figure*}
\centering
    \includegraphics[width=6.0in]{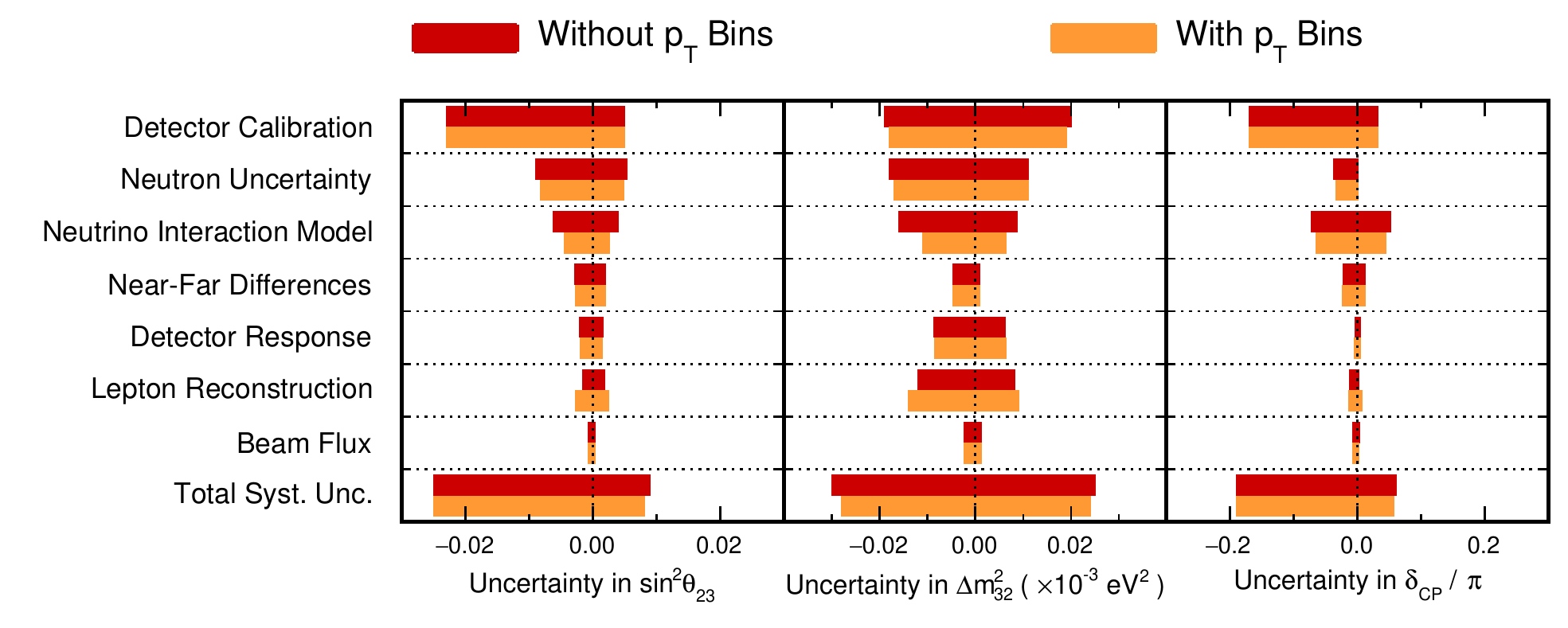}
    \caption{Systematic uncertainties on \snsq, \dmsq, and \deltacp evaluated at the best-fit point.  Detector calibration uncertainties, which are less constrained by extrapolation, are dominant for all three oscillation parameters.  Uncertainties for extrapolation with (orange) and without (red) \pt bins are shown for comparison. The statistical uncertainties (not shown) are [-0.033, 0.022] for   $\sin^2\theta_{23}$, [-0.055, 0.043] ($\times 10^{-3}$~eV$^2$) for $\dmsq$, and [-0.87, 0.21] for $\deltacp$ .
    }
\label{fig:syst}
\end{figure*}

\section{\label{sec:systematics} Systematic Uncertainties} 

The impacts of systematic uncertainties are evaluated by varying the simulation via event reweighting or simulating alternative event samples and repeating the extrapolation procedure. 
Uncertainties associated with the neutrino flux, neutron modeling, and detector calibrations are unchanged from the previous analysis~\cite{Acero:2019ksn}.

Detector calibration uncertainties remain dominant and are driven by a 5\% uncertainty in the calorimetric energy scale. Additionally, a new time-dependent calibration uncertainty is included to account for any residual differences remaining after performing the calibration over shorter time periods as mentioned previously.

Neutrino interaction model uncertainties are evaluated using the event reweighting framework in \textsc{genie} with additional uncertainties constructed by NOvA as follows. Uncertainties on CCQE RPA, low-\qsq RES suppression, 2p2h, and non-resonant and incoherent N$\pi$ production are established for the new model set using methods similar to those in Ref.~\cite{Acero:2020eit}. Pion FSI uncertainties are based on comparisons to $\pi^{+}$ on ${^{12}\text{C}}$ scattering data~\cite{Allardyce:1973ce,Saunders:1996ic,Meirav:1988pn,Levenson:1983xu,Ashery:1981tq,Ashery:1984ne,PinzonGuerra:2016uae}
and prior studies using an alternative neutrino interaction generator~\cite{PinzonGuerra:2018rju}.  Uncertainties on the \nueParen CC cross section relative to the \numuParen CC cross section due to radiative corrections and possible second-class currents are unchanged from previous analyses~\cite{NOvA:2018gge}.

As in the previous analysis, uncertainties are included that are detector specific or account for differences between the ND and FD: the detector masses, beam exposures, kinematic acceptances, beam-induced pile-up, \nue CC selection in the ND, and cosmogenic backgrounds in the FD. The improved hit clustering algorithm reduces pile-up effects in the ND, decreasing uncertainties for the associated data-MC selection efficiency differences. An uncertainty for kinematic acceptance differences between the detectors was overestimated in the previous analysis and is subdominant in this analysis after correction. Extrapolating in \pt bins would have substantially reduced the effect of this uncertainty even if left uncorrected.

Uncertainties arising from the custom light model are assigned based on comparison to a more robust response model that was not fully incorporated into the simulation for this analysis. 
This model is constrained by a sample of ND proton candidates in addition to the muon sample used previously.
Differences in the detector response between the proton and muon samples also provide a data-driven uncertainty on the relative production of Cherenkov and scintillation light in the model. 

Quantities affected by lepton reconstruction uncertainties include the muon energy scale and lepton angle.
The muon energy scale uncertainty now includes a detector mass uncertainty with a component that is uncorrelated between the detectors, plus a correlated component accounting for the Fermi density effect and muon range differences across models. Extrapolating in \pt bins introduces a dependence on the reconstructed lepton angle for which a \SI{10}{\milli\radian} uncorrelated uncertainty is applied.

Figure~\ref{fig:syst} shows the impact of the systematic uncertainties on the measurement of  \snsq, \dmsq, and \deltacp as evaluated at the determined best-fit point.  The extrapolation method significantly reduces the impact of the detector correlated beam flux and neutrino interaction model uncertainties.  In contrast, energy calibration and uncorrelated uncertainties that reflect ND-FD differences are less constrained by extrapolation.  Figure~\ref{fig:syst} also shows the impact of uncertainties for extrapolation with and without \pt bins.  Extrapolating in \pt bins reduces the interaction model uncertainty by 10-30\%, and the total systematic uncertainty by up to 9\%.  Detector calibration, detector response, and neutron modeling uncertainties that affect the reconstructed energy of the recoiling hadronic system, which is correlated with \pt, are more modestly reduced.  The  extrapolation in bins of \pt depends on reconstructed lepton kinematics and results in a marginal increase in the associated uncertainties.

\section{\label{sec:results} Results}

\nolinenumbers
\begin{figure}[b]
    \makebox[\columnwidth][c]{
        \subfloat[]{\label{fig:fd-numu-data-figure}\includegraphics[width=1.751in]{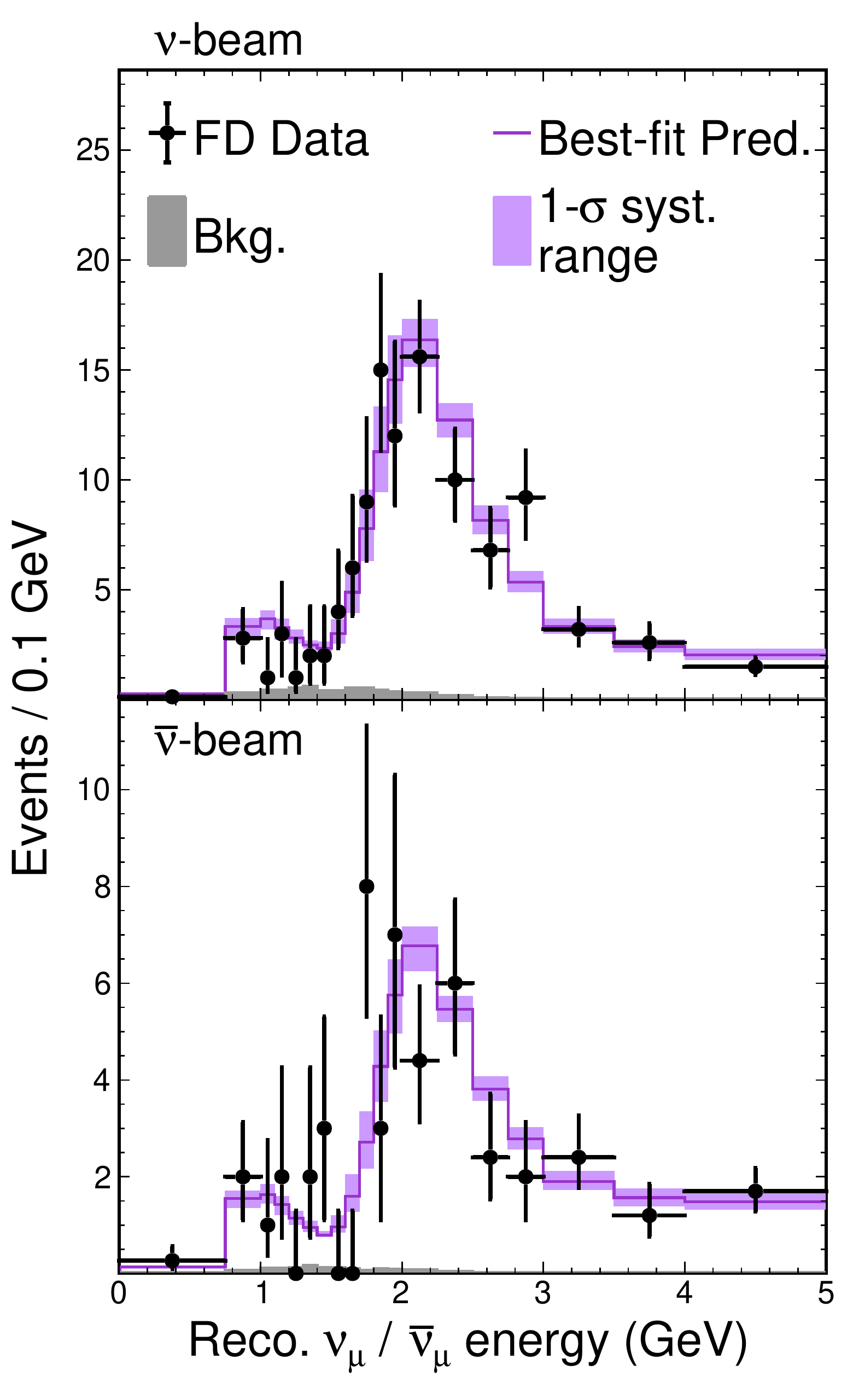}}
        \subfloat[]{\label{fig:fd-nue-data-figure}\includegraphics[width=1.751in]{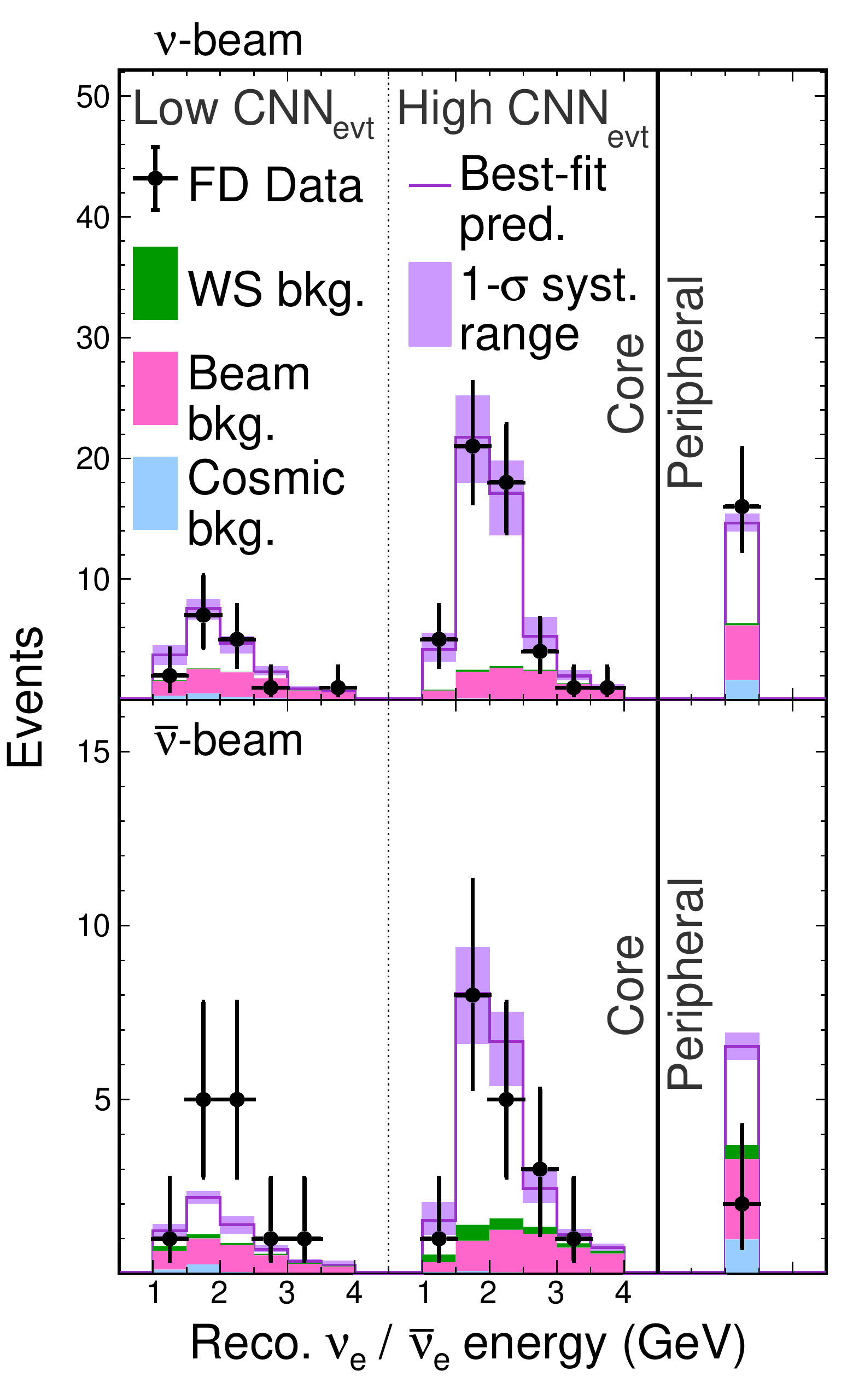}}
    }
    \caption{
        Reconstructed neutrino energy spectra for the FD (a) \numucc and (b) \nuecc samples with the neutrino-mode beam on top and antineutrino-mode on the bottom~\cite{supplemental-numu}. The \numucc \efrac sub-samples have been combined. The \nuecc low and high \evtcnn, and peripheral sub-samples are shown. 
    }
    \label{fig:fd-data-figure}
\end{figure}

The extrapolated predictions of the FD spectra are recomputed for varying oscillation parameters and compared to data using a Poisson negative log-likelihood ratio,~\LL. The best-fit parameters minimize~\LL.
The following solar and reactor neutrino experiment constraints are used: $\Delta m^2_{21} = 7.53  \times 10^{-5}~{\rm eV}^2$, $\sin^2\theta_{12} =0.307 $, and $\sin^2\theta_{13} = 0.0210 \pm 0.0011$ \cite{PhysRevD.98.030001}.
The parameters \dmsq, \snsq, and \deltacp 
are varied without constraints while the 64 systematic uncertainties are assigned penalty terms equal to the square of the number of standard deviations by which they vary from their nominal values. The value of $\sin^2\theta_{13}$ is allowed to float similarly.    
Feldman-Cousins' unified approach~\cite{Feldman:1997qc,Sousa:2019nhd} is used to determine the confidence intervals for the oscillation parameters.  All significances given,  or  plotted,  are FC-corrected values. The fitted parameters not shown are profiled over. 

\begin{table}[b]
\caption{
Event counts at the FD, both observed and predicted at the best-fit point (see Table~\ref{tb:oscillation_parameters_table}).}
\label{tb:event-count-table}
\begin{tabular}{lcccc}
	\hline
	\hline
                                    & \multicolumn{2}{c}{Neutrino beam} & \multicolumn{2}{c}{Antineutrino beam} \\
                                    & {\numucc}             & {\nuecc}              & {\numubarcc}          & {\nuebarcc} \\
    \hline
    \numu{}\,$\rightarrow$\,\numu       &  201.1                &  1.7                  & 26.0                  & 0.2  \\
    \numubar{}\,$\rightarrow$\,\numubar &  12.6                 &  0.0                  & 77.2                  & 0.2  \\
    \numu{}\,$\rightarrow$\,\nue        &   0.1                 & 59.0                  &  0.0                  & 2.3  \\
    \numubar{}\,$\rightarrow$\,\nuebar  &   0.0                 &  1.0                  &  0.0                  & 19.2 \\
    Beam \nue$+$\xspace\nuebar      &   0.0                 & 14.1                  &  0.0                  & 7.3  \\
    NC                              &   2.6                 &  6.3                  &  0.8                  & 2.2  \\
    Cosmic                          &   5.0                 &  3.1                  &  0.9                  & 1.6  \\
    Others                          &   0.9                 &  0.5                  &  0.4                  & 0.3  \\
    \hline
    \rule[-4pt]{0pt}{13pt}Signal                          & 214.1$^{+14.4}_{-14.0}$ & 59.0$^{+2.5}_{-2.5}$  & 103.4$^{+7.1}_{-7.0}$ & 19.2$^{+0.6}_{-0.7}$ \\
    \rule[-4pt]{0pt}{13pt}Background                      &   8.2$^{+1.9}_{-1.7}$ & 26.8$^{+1.6}_{-1.7}$  &   2.1$^{+0.7}_{-0.7}$ & 14.0$^{+0.9}_{-1.0}$  \\
    \hline
    \rule{0pt}{9pt}Best fit                        & 222.3                 & 85.8                  & 105.4                 & 33.2 \\
    Observed                        & 211                   & 82                    & 105                   & 33   \\
    \hline
    \hline
    \end{tabular}
\end{table}

Figure~\ref{fig:fd-data-figure} shows the energy spectra of the \numucc, \numubarcc, \nuecc, and \nuebarcc candidates recorded at the FD. The distributions are compared to the oscillation best-fit expectations. Table~\ref{tb:event-count-table} summarizes the total event counts and estimated compositions of the selected samples. The CC candidate event samples recorded at the FD include 211 (105) observed \numuParen{}\,$\rightarrow$\,\numuParen events and 82 (33) \numuParen{}\,$\rightarrow$\,\nueParen candidate events. The latter \nueParen appearance sample has an estimated background of $26.8^{+1.6}_{-1.7}$ ($14.0^{+0.9}_{-1.0}$). 

\begin{table}[t]
    \caption{
      Summary of oscillation parameter best-fit results for different choices of the mass ordering (Normal or Inverted) and upper or lower $\theta_{23}$ octant (UO, LO), along with the FC corrected significance (in units of $\sigma$) at which those combinations are disfavored. Full uncertainties are given in \cite{supplemental-tab}. 
    }
        
    \label{tb:oscillation_parameters_table}
    \begin{tabular}{ccccc}
        \hline \hline
            \rule{0pt}{8pt}\multirow{2}{*}{Parameter}        & \multicolumn{2}{c}{Normal ord.}                  & \multicolumn{2}{c}{Inverted ord.} \\
                                                                       & UO                      & LO            & UO            & LO \\
        \hline 
            \rule[-4pt]{0pt}{13pt}$\Delta m^2_{32} (10^{-3}~{\rm eV}^2$) & $+2.41\pm{0.07}$ & $+2.39$       & $-2.45$       & $-2.44$ \\
            \rule[-4pt]{0pt}{13pt}$\sin^2 \theta_{23}$                   & $0.57^{+0.03}_{-0.04}$  & $0.46$        & $0.56$        & $0.46$ \\
            \rule[-4pt]{0pt}{13pt}$\delta_{\rm CP} (\pi)$                & $0.82^{+0.27}_{-0.87}$  & $0.07$        & $1.52$        & $1.41$ \\
        \hline 
                            Rejection significance                     & -                       & $1.1\,\sigma$ & $0.9\,\sigma$ & $1.1\,\sigma$ \\
        \hline \hline
    \end{tabular}
\end{table}

This analysis determines a best-fit in the normal mass ordering and upper $\theta_{23}$ octant (significance of 1.0\,$\sigma$ and 1.2\,$\sigma$, respectively), where \LL~=~173.55 for 175 degrees of freedom (p-value of 0.705).  The data disfavor combinations that lead to a strong asymmetry in the rate of \nue versus \nuebar appearance; therefore, the inverted mass ordering with $\deltacp = \pi/2$  is excluded at more than 3\,$\sigma$ and the normal mass ordering with $\deltacp = 3\pi/2$  is disfavored at 2\,$\sigma$ confidence. However, owing to the degeneracies, the 90\% confidence level allowed regions cover all values of \deltacp given permutations of mass ordering and octant.  Thus, the current data do not exhibit a preference concerning CP conservation versus violation. Table~\ref{tb:oscillation_parameters_table} shows the best-fit parameter values for each choice of $\theta_{23}$ octant and mass ordering.  

\begin{figure}[t]
    \includegraphics[width=3.410in]{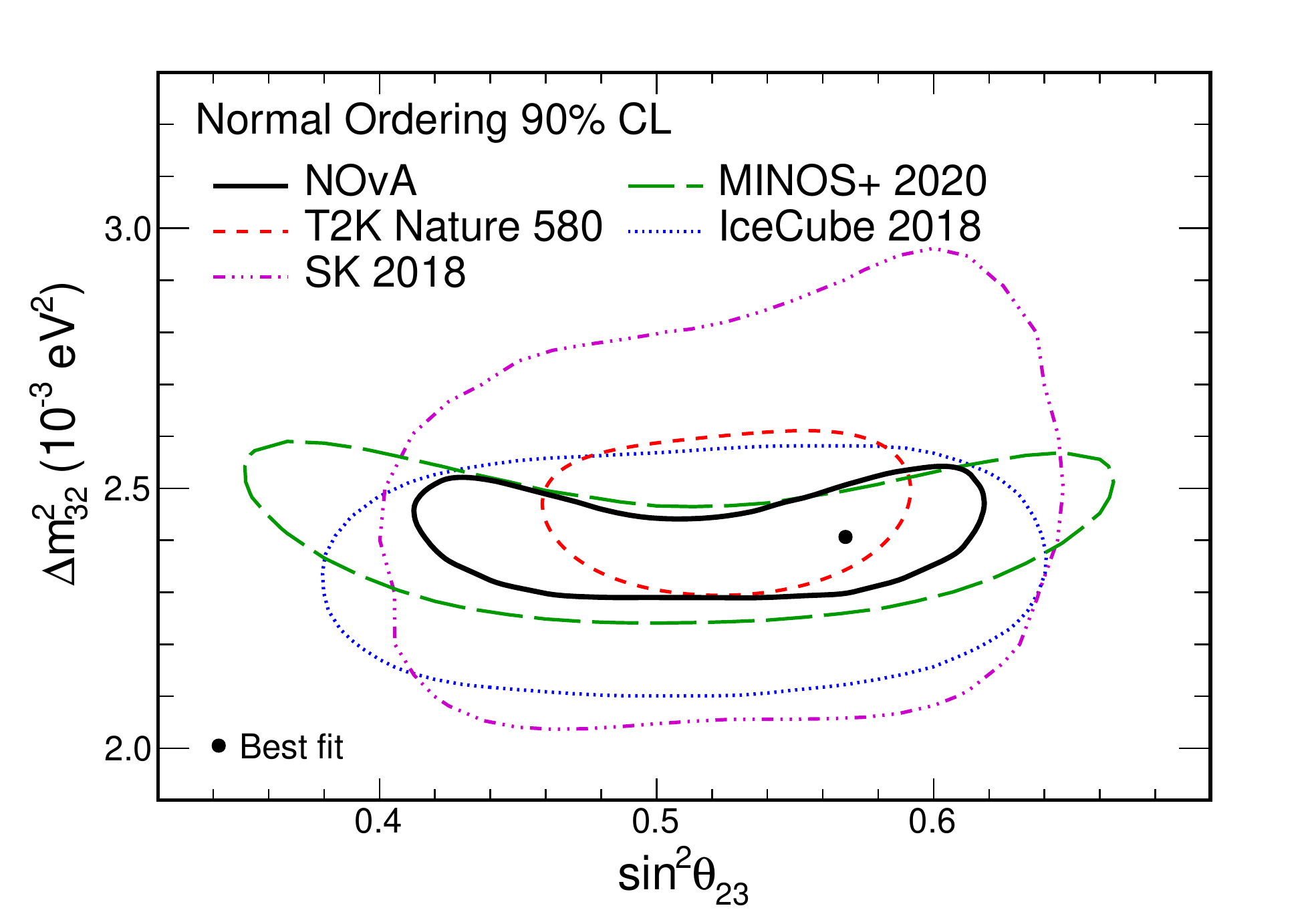}
    \caption{The 90\% confidence level region for \dmsq and \snsq,
    with the FC corrected allowed region and best-fit point for NOvA~\cite{supplemental-contours} superposed on contours from other experiments.~\cite{Abe:2019vii,Abe:2017aap,MINOS:2020llm,Aartsen:2017nmd}$^3$.
}
\label{fig:dms32-sstt23-contour}
\end{figure}

\begin{figure}[t]

    \includegraphics[width=3.410in]{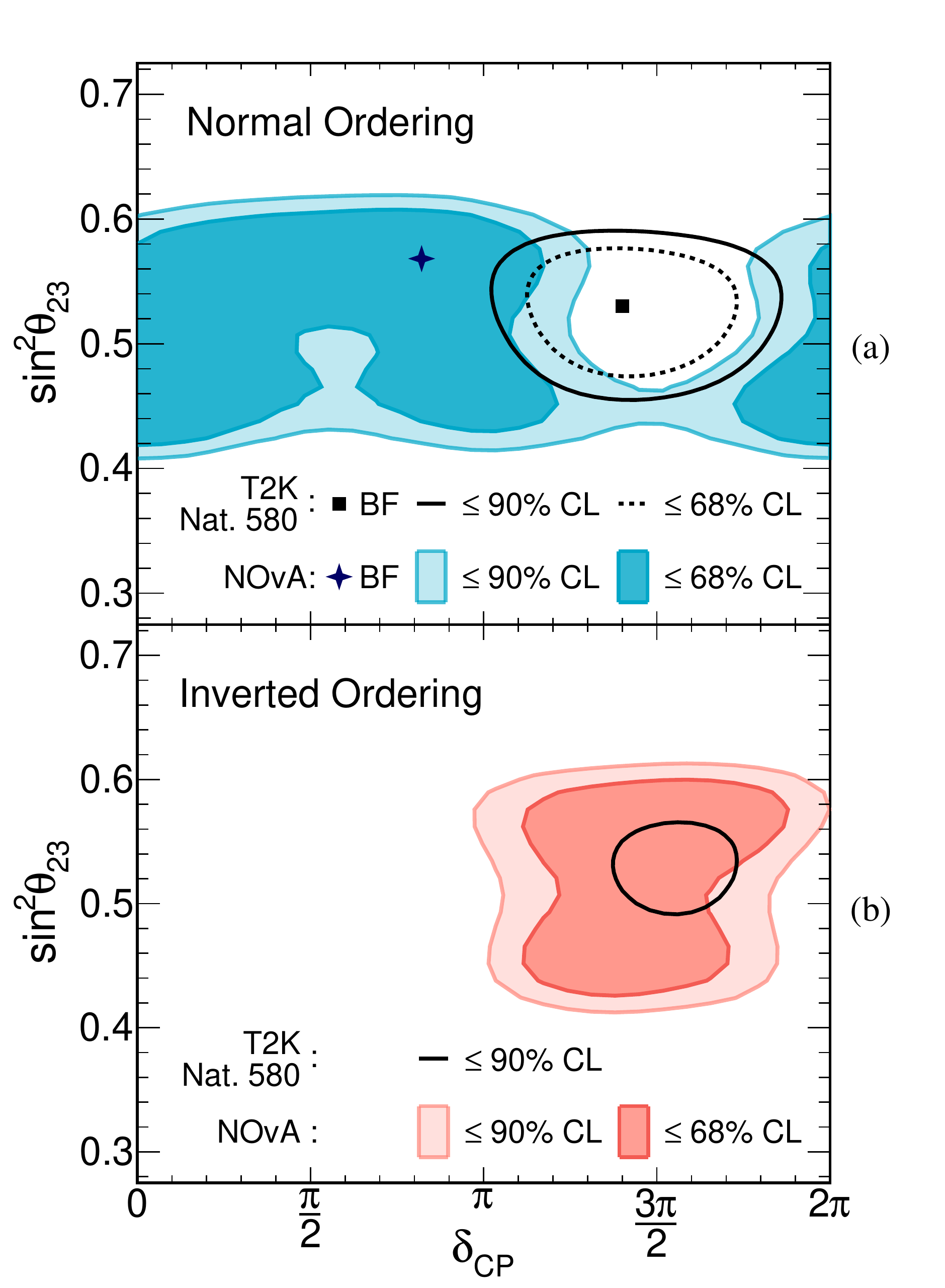}

    \caption{
    The 68\% and 90\% confidence level contours in \snsq vs. \deltacp in the (a) normal mass ordering and (b) 
    inverted mass ordering ~\cite{supplemental-dcp}. 
    The cross denotes the NOvA best-fit point and colored areas depict the 90\% and 68\% FC corrected allowed regions for NOvA. Overlaid black solid-line and dashed-line contours depict allowed regions reported by T2K~\cite{Abe:2019vii}$^3$. 
    \label{fig:sstt23-dcp-islands}}
\end{figure}

Figure~\ref{fig:dms32-sstt23-contour} compares the 90\% confidence level contours for $\dmsq$ and $\snsq$ with those of other experiments~\cite{Abe:2019vii,Abe:2017aap,MINOS:2020llm,Aartsen:2017nmd}$^3$. Allowed regions in \snsq and \deltacp are shown in Fig.~\ref{fig:sstt23-dcp-islands} and are compared with a recent best fit from T2K \cite{Abe:2019vii}\footnote{While this paper was in its final internal review, an updated analysis was published by the T2K collaboration~\cite{T2K:2021xwb}. Compared to Ref.~\cite{Abe:2019vii}, the dataset remains unchanged and the same approach is used. The conclusions drawn from the comparisons of the contours remains unchanged.}.

As shown in Fig.~\ref{fig:sstt23-dcp-islands}a, the T2K best-fit point is in the NO but lies in a region that NOvA disfavors. However, some regions of overlap remain.
Figure~\ref{fig:sstt23-dcp-islands}b shows that for IO, the T2K allowed region at 90\% confidence level is entirely contained within the corresponding NOvA allowed region. This outcome reflects in part the circumstance that T2K observes a relatively more pronounced asymmetry in \nue versus \nuebar oscillations.

Although each experiment reports a mild preference for NO, it has been suggested that a joint fit of the two experiments might converge on an IO solution~\cite{Kelly:2020fkv}. Some authors have also explored the possibility that the differences in the \numu{}\,$\rightarrow$\,\nue and \numubar{}\,$\rightarrow$\,\nuebar rates seen by the experiments are explained by additional non-standard matter effects~\cite{Denton:2020uda,Chatterjee:2020kkm}.

\setlength{\parskip}{0pt}

In conclusion, we have presented improved measurements of oscillation parameters \dmsq, \snsq, and \deltacp, including an expanded data set and enhanced analysis techniques with respect to previous publications. These measurements continue to favor the normal mass ordering and upper octant of \snsq, as well as values of the oscillation parameters that do not lead to a large asymmetry in \numu{}\,$\rightarrow$\,\nue and \numubar{}\,$\rightarrow$\,\nuebar oscillation rates.

\section{\label{sec:acknowledgments} Acknowledgments}

This document was prepared by the NOvA collaboration using the resources of the Fermi National Accelerator Laboratory (Fermilab), a U.S. Department of Energy, Office of Science, HEP User Facility. Fermilab is managed by Fermi Research Alliance, LLC (FRA), acting under Contract No. DE-AC02-07CH11359. This work was supported by the U.S. Department of Energy; the U.S. National Science Foundation; the Department of Science and Technology, India; the European Research Council; the MSMT CR, GA UK, Czech Republic; the RAS, RFBR, RMES, RSF, and BASIS Foundation, Russia; CNPq and FAPEG, Brazil; STFC, UKRI, and the Royal Society, United Kingdom; and the State and University of Minnesota.  We are grateful for the contributions of the staffs of the University of Minnesota at the Ash River Laboratory and of Fermilab.

\bibliographystyle{apsrev4-2}

\bibliography{refs}

\begin{thebibliography}{99}%
\makeatletter
\providecommand \@ifxundefined [1]{%
 \@ifx{#1\undefined}
}%
\providecommand \@ifnum [1]{%
 \ifnum #1\expandafter \@firstoftwo
 \else \expandafter \@secondoftwo
 \fi
}%
\providecommand \@ifx [1]{%
 \ifx #1\expandafter \@firstoftwo
 \else \expandafter \@secondoftwo
 \fi
}%
\providecommand \natexlab [1]{#1}%
\providecommand \enquote  [1]{``#1''}%
\providecommand \bibnamefont  [1]{#1}%
\providecommand \bibfnamefont [1]{#1}%
\providecommand \citenamefont [1]{#1}%
\providecommand \href@noop [0]{\@secondoftwo}%
\providecommand \href [0]{\begingroup \@sanitize@url \@href}%
\providecommand \@href[1]{\@@startlink{#1}\@@href}%
\providecommand \@@href[1]{\endgroup#1\@@endlink}%
\providecommand \@sanitize@url [0]{\catcode `\\12\catcode `\$12\catcode
  `\&12\catcode `\#12\catcode `\^12\catcode `\_12\catcode `\%12\relax}%
\providecommand \@@startlink[1]{}%
\providecommand \@@endlink[0]{}%
\providecommand \url  [0]{\begingroup\@sanitize@url \@url }%
\providecommand \@url [1]{\endgroup\@href {#1}{\urlprefix }}%
\providecommand \urlprefix  [0]{URL }%
\providecommand \Eprint [0]{\href }%
\providecommand \doibase [0]{https://doi.org/}%
\providecommand \selectlanguage [0]{\@gobble}%
\providecommand \bibinfo  [0]{\@secondoftwo}%
\providecommand \bibfield  [0]{\@secondoftwo}%
\providecommand \translation [1]{[#1]}%
\providecommand \BibitemOpen [0]{}%
\providecommand \bibitemStop [0]{}%
\providecommand \bibitemNoStop [0]{.\EOS\space}%
\providecommand \EOS [0]{\spacefactor3000\relax}%
\providecommand \BibitemShut  [1]{\csname bibitem#1\endcsname}%
\let\auto@bib@innerbib\@empty
\bibitem [{\citenamefont {Acero}\ \emph {et~al.}(2019)\citenamefont {Acero}
  \emph {et~al.}}]{Acero:2019ksn}%
  \BibitemOpen
  \bibfield  {author} {\bibinfo {author} {\bibfnamefont {M.~A.}\ \bibnamefont
  {Acero}} \emph {et~al.} (\bibinfo {collaboration} {NOvA}),\ }\bibfield
  {title} {\bibinfo {title} {{First Measurement of Neutrino Oscillation
  Parameters using Neutrinos and Antineutrinos by NOvA}},\ }\href
  {https://doi.org/10.1103/PhysRevLett.123.151803} {\bibfield  {journal}
  {\bibinfo  {journal} {Phys. Rev. Lett.}\ }\textbf {\bibinfo {volume} {123}},\
  \bibinfo {pages} {151803} (\bibinfo {year} {2019})},\ \Eprint
  {https://arxiv.org/abs/1906.04907} {arXiv:1906.04907 [hep-ex]} \BibitemShut
  {NoStop}%
\bibitem [{\citenamefont {Fukuda}\ \emph {et~al.}(1998)\citenamefont {Fukuda}
  \emph {et~al.}}]{Fukuda:1998mi}%
  \BibitemOpen
  \bibfield  {author} {\bibinfo {author} {\bibfnamefont {Y.}~\bibnamefont
  {Fukuda}} \emph {et~al.} (\bibinfo {collaboration} {Super-Kamiokande}),\
  }\bibfield  {title} {\bibinfo {title} {{Evidence for oscillation of
  atmospheric neutrinos}},\ }\href
  {https://doi.org/10.1103/PhysRevLett.81.1562} {\bibfield  {journal} {\bibinfo
   {journal} {Phys. Rev. Lett.}\ }\textbf {\bibinfo {volume} {81}},\ \bibinfo
  {pages} {1562} (\bibinfo {year} {1998})},\ \Eprint
  {https://arxiv.org/abs/hep-ex/9807003} {arXiv:hep-ex/9807003 [hep-ex]}
  \BibitemShut {NoStop}%
\bibitem [{\citenamefont {Fukuda}\ \emph {et~al.}(2002)\citenamefont {Fukuda}
  \emph {et~al.}}]{Fukuda:2002pe}%
  \BibitemOpen
  \bibfield  {author} {\bibinfo {author} {\bibfnamefont {S.}~\bibnamefont
  {Fukuda}} \emph {et~al.} (\bibinfo {collaboration} {Super-Kamiokande}),\
  }\bibfield  {title} {\bibinfo {title} {{Determination of solar neutrino
  oscillation parameters using 1496 days of Super-Kamiokande I data}},\ }\href
  {https://doi.org/10.1016/S0370-2693(02)02090-7} {\bibfield  {journal}
  {\bibinfo  {journal} {Phys. Lett. B}\ }\textbf {\bibinfo {volume} {539}},\
  \bibinfo {pages} {179} (\bibinfo {year} {2002})},\ \Eprint
  {https://arxiv.org/abs/hep-ex/0205075} {arXiv:hep-ex/0205075 [hep-ex]}
  \BibitemShut {NoStop}%
\bibitem [{\citenamefont {Ahmad}\ \emph {et~al.}(2002)\citenamefont {Ahmad}
  \emph {et~al.}}]{Ahmad:2002jz}%
  \BibitemOpen
  \bibfield  {author} {\bibinfo {author} {\bibfnamefont {Q.~R.}\ \bibnamefont
  {Ahmad}} \emph {et~al.} (\bibinfo {collaboration} {SNO}),\ }\bibfield
  {title} {\bibinfo {title} {{Direct evidence for neutrino flavor
  transformation from neutral current interactions in the Sudbury Neutrino
  Observatory}},\ }\href {https://doi.org/10.1103/PhysRevLett.89.011301}
  {\bibfield  {journal} {\bibinfo  {journal} {Phys. Rev. Lett.}\ }\textbf
  {\bibinfo {volume} {89}},\ \bibinfo {pages} {011301} (\bibinfo {year}
  {2002})},\ \Eprint {https://arxiv.org/abs/nucl-ex/0204008}
  {arXiv:nucl-ex/0204008 [nucl-ex]} \BibitemShut {NoStop}%
\bibitem [{\citenamefont {Eguchi}\ \emph {et~al.}(2003)\citenamefont {Eguchi}
  \emph {et~al.}}]{Eguchi:2002dm}%
  \BibitemOpen
  \bibfield  {author} {\bibinfo {author} {\bibfnamefont {K.}~\bibnamefont
  {Eguchi}} \emph {et~al.} (\bibinfo {collaboration} {KamLAND}),\ }\bibfield
  {title} {\bibinfo {title} {{First results from KamLAND: Evidence for reactor
  anti-neutrino disappearance}},\ }\href
  {https://doi.org/10.1103/PhysRevLett.90.021802} {\bibfield  {journal}
  {\bibinfo  {journal} {Phys. Rev. Lett.}\ }\textbf {\bibinfo {volume} {90}},\
  \bibinfo {pages} {021802} (\bibinfo {year} {2003})},\ \Eprint
  {https://arxiv.org/abs/hep-ex/0212021} {arXiv:hep-ex/0212021 [hep-ex]}
  \BibitemShut {NoStop}%
\bibitem [{\citenamefont {Michael}\ \emph {et~al.}(2006)\citenamefont {Michael}
  \emph {et~al.}}]{Michael:2006rx}%
  \BibitemOpen
  \bibfield  {author} {\bibinfo {author} {\bibfnamefont {D.~G.}\ \bibnamefont
  {Michael}} \emph {et~al.} (\bibinfo {collaboration} {MINOS}),\ }\bibfield
  {title} {\bibinfo {title} {{Observation of muon neutrino disappearance with
  the MINOS detectors and the NuMI neutrino beam}},\ }\href
  {https://doi.org/10.1103/PhysRevLett.97.191801} {\bibfield  {journal}
  {\bibinfo  {journal} {Phys. Rev. Lett.}\ }\textbf {\bibinfo {volume} {97}},\
  \bibinfo {pages} {191801} (\bibinfo {year} {2006})},\ \Eprint
  {https://arxiv.org/abs/hep-ex/0607088} {arXiv:hep-ex/0607088 [hep-ex]}
  \BibitemShut {NoStop}%
\bibitem [{\citenamefont {Abe}\ \emph {et~al.}(2011)\citenamefont {Abe} \emph
  {et~al.}}]{Abe:2011sj}%
  \BibitemOpen
  \bibfield  {author} {\bibinfo {author} {\bibfnamefont {K.}~\bibnamefont
  {Abe}} \emph {et~al.} (\bibinfo {collaboration} {T2K}),\ }\bibfield  {title}
  {\bibinfo {title} {{Indication of Electron Neutrino Appearance from an
  Accelerator-produced Off-axis Muon Neutrino Beam}},\ }\href
  {https://doi.org/10.1103/PhysRevLett.107.041801} {\bibfield  {journal}
  {\bibinfo  {journal} {Phys. Rev. Lett.}\ }\textbf {\bibinfo {volume} {107}},\
  \bibinfo {pages} {041801} (\bibinfo {year} {2011})},\ \Eprint
  {https://arxiv.org/abs/1106.2822} {arXiv:1106.2822 [hep-ex]} \BibitemShut
  {NoStop}%
\bibitem [{\citenamefont {Abe}\ \emph {et~al.}(2012)\citenamefont {Abe} \emph
  {et~al.}}]{Abe:2011fz}%
  \BibitemOpen
  \bibfield  {author} {\bibinfo {author} {\bibfnamefont {Y.}~\bibnamefont
  {Abe}} \emph {et~al.} (\bibinfo {collaboration} {Double Chooz}),\ }\bibfield
  {title} {\bibinfo {title} {{Indication of Reactor $\bar{\nu}_e$ Disappearance
  in the Double Chooz Experiment}},\ }\href
  {https://doi.org/10.1103/PhysRevLett.108.131801} {\bibfield  {journal}
  {\bibinfo  {journal} {Phys. Rev. Lett.}\ }\textbf {\bibinfo {volume} {108}},\
  \bibinfo {pages} {131801} (\bibinfo {year} {2012})},\ \Eprint
  {https://arxiv.org/abs/1112.6353} {arXiv:1112.6353 [hep-ex]} \BibitemShut
  {NoStop}%
\bibitem [{\citenamefont {An}\ \emph {et~al.}(2012)\citenamefont {An} \emph
  {et~al.}}]{An:2012eh}%
  \BibitemOpen
  \bibfield  {author} {\bibinfo {author} {\bibfnamefont {F.~P.}\ \bibnamefont
  {An}} \emph {et~al.} (\bibinfo {collaboration} {Daya Bay}),\ }\bibfield
  {title} {\bibinfo {title} {{Observation of electron-antineutrino
  disappearance at Daya Bay}},\ }\href
  {https://doi.org/10.1103/PhysRevLett.108.171803} {\bibfield  {journal}
  {\bibinfo  {journal} {Phys. Rev. Lett.}\ }\textbf {\bibinfo {volume} {108}},\
  \bibinfo {pages} {171803} (\bibinfo {year} {2012})},\ \Eprint
  {https://arxiv.org/abs/1203.1669} {arXiv:1203.1669 [hep-ex]} \BibitemShut
  {NoStop}%
\bibitem [{\citenamefont {Ahn}\ \emph {et~al.}(2012)\citenamefont {Ahn} \emph
  {et~al.}}]{Ahn:2012nd}%
  \BibitemOpen
  \bibfield  {author} {\bibinfo {author} {\bibfnamefont {J.~K.}\ \bibnamefont
  {Ahn}} \emph {et~al.} (\bibinfo {collaboration} {RENO}),\ }\bibfield  {title}
  {\bibinfo {title} {{Observation of Reactor Electron Antineutrino
  Disappearance in the RENO Experiment}},\ }\href
  {https://doi.org/10.1103/PhysRevLett.108.191802} {\bibfield  {journal}
  {\bibinfo  {journal} {Phys. Rev. Lett.}\ }\textbf {\bibinfo {volume} {108}},\
  \bibinfo {pages} {191802} (\bibinfo {year} {2012})},\ \Eprint
  {https://arxiv.org/abs/1204.0626} {arXiv:1204.0626 [hep-ex]} \BibitemShut
  {NoStop}%
\bibitem [{\citenamefont {Mohapatra}\ and\ \citenamefont
  {Smirnov}(2006)}]{Mohapatra:2006gs}%
  \BibitemOpen
  \bibfield  {author} {\bibinfo {author} {\bibfnamefont {R.~N.}\ \bibnamefont
  {Mohapatra}}\ and\ \bibinfo {author} {\bibfnamefont {A.~Y.}\ \bibnamefont
  {Smirnov}},\ }\bibfield  {title} {\bibinfo {title} {{Neutrino Mass and New
  Physics}},\ }\bibfield  {booktitle} {\emph {\bibinfo {booktitle} {{Elementary
  particle physics. Proceedings, Corfu Summer Institute, CORFU2005, Corfu,
  Greece, September 4-26, 2005}}},\ }\href
  {https://doi.org/10.1146/annurev.nucl.56.080805.140534} {\bibfield  {journal}
  {\bibinfo  {journal} {Ann. Rev. Nucl. Part. Sci.}\ }\textbf {\bibinfo
  {volume} {56}},\ \bibinfo {pages} {569} (\bibinfo {year} {2006})},\ \Eprint
  {https://arxiv.org/abs/hep-ph/0603118} {arXiv:hep-ph/0603118 [hep-ph]}
  \BibitemShut {NoStop}%
\bibitem [{\citenamefont {Nunokawa}\ \emph {et~al.}(2008)\citenamefont
  {Nunokawa}, \citenamefont {Parke},\ and\ \citenamefont
  {Valle}}]{Nunokawa:2007qh}%
  \BibitemOpen
  \bibfield  {author} {\bibinfo {author} {\bibfnamefont {H.}~\bibnamefont
  {Nunokawa}}, \bibinfo {author} {\bibfnamefont {S.~J.}\ \bibnamefont
  {Parke}},\ and\ \bibinfo {author} {\bibfnamefont {J.~W.~F.}\ \bibnamefont
  {Valle}},\ }\bibfield  {title} {\bibinfo {title} {{CP Violation and Neutrino
  Oscillations}},\ }\href {https://doi.org/10.1016/j.ppnp.2007.10.001}
  {\bibfield  {journal} {\bibinfo  {journal} {Prog. Part. Nucl. Phys.}\
  }\textbf {\bibinfo {volume} {60}},\ \bibinfo {pages} {338} (\bibinfo {year}
  {2008})},\ \Eprint {https://arxiv.org/abs/0710.0554} {arXiv:0710.0554
  [hep-ph]} \BibitemShut {NoStop}%
\bibitem [{\citenamefont {Altarelli}\ and\ \citenamefont
  {Feruglio}(2010)}]{Altarelli:2010gt}%
  \BibitemOpen
  \bibfield  {author} {\bibinfo {author} {\bibfnamefont {G.}~\bibnamefont
  {Altarelli}}\ and\ \bibinfo {author} {\bibfnamefont {F.}~\bibnamefont
  {Feruglio}},\ }\bibfield  {title} {\bibinfo {title} {{Discrete Flavor
  Symmetries and Models of Neutrino Mixing}},\ }\href
  {https://doi.org/10.1103/RevModPhys.82.2701} {\bibfield  {journal} {\bibinfo
  {journal} {Rev. Mod. Phys.}\ }\textbf {\bibinfo {volume} {82}},\ \bibinfo
  {pages} {2701} (\bibinfo {year} {2010})},\ \Eprint
  {https://arxiv.org/abs/1002.0211} {arXiv:1002.0211 [hep-ph]} \BibitemShut
  {NoStop}%
\bibitem [{\citenamefont {King}(2015)}]{King:2015aea}%
  \BibitemOpen
  \bibfield  {author} {\bibinfo {author} {\bibfnamefont {S.~F.}\ \bibnamefont
  {King}},\ }\bibfield  {title} {\bibinfo {title} {{Models of Neutrino Mass,
  Mixing and CP Violation}},\ }\href
  {https://doi.org/10.1088/0954-3899/42/12/123001} {\bibfield  {journal}
  {\bibinfo  {journal} {J. Phys. G}\ }\textbf {\bibinfo {volume} {42}},\
  \bibinfo {pages} {123001} (\bibinfo {year} {2015})},\ \Eprint
  {https://arxiv.org/abs/1510.02091} {arXiv:1510.02091 [hep-ph]} \BibitemShut
  {NoStop}%
\bibitem [{\citenamefont {Petcov}(2018)}]{Petcov:2017ggy}%
  \BibitemOpen
  \bibfield  {author} {\bibinfo {author} {\bibfnamefont {S.~T.}\ \bibnamefont
  {Petcov}},\ }\bibfield  {title} {\bibinfo {title} {{Discrete Flavour
  Symmetries, Neutrino Mixing and Leptonic CP Violation}},\ }\href
  {https://doi.org/10.1140/epjc/s10052-018-6158-5} {\bibfield  {journal}
  {\bibinfo  {journal} {Eur. Phys. J. C}\ }\textbf {\bibinfo {volume} {78}},\
  \bibinfo {pages} {709} (\bibinfo {year} {2018})},\ \Eprint
  {https://arxiv.org/abs/1711.10806} {arXiv:1711.10806 [hep-ph]} \BibitemShut
  {NoStop}%
\bibitem [{\citenamefont {Pascoli}\ and\ \citenamefont
  {Petcov}(2002)}]{Pascoli:2002xq}%
  \BibitemOpen
  \bibfield  {author} {\bibinfo {author} {\bibfnamefont {S.}~\bibnamefont
  {Pascoli}}\ and\ \bibinfo {author} {\bibfnamefont {S.~T.}\ \bibnamefont
  {Petcov}},\ }\bibfield  {title} {\bibinfo {title} {{The SNO solar neutrino
  data, neutrinoless double beta decay and neutrino mass spectrum}},\ }\href
  {https://doi.org/10.1016/S0370-2693(02)02510-8} {\bibfield  {journal}
  {\bibinfo  {journal} {Phys. Lett. B}\ }\textbf {\bibinfo {volume} {544}},\
  \bibinfo {pages} {239} (\bibinfo {year} {2002})},\ \Eprint
  {https://arxiv.org/abs/hep-ph/0205022} {arXiv:hep-ph/0205022 [hep-ph]}
  \BibitemShut {NoStop}%
\bibitem [{\citenamefont {Bahcall}\ \emph {et~al.}(2004)\citenamefont
  {Bahcall}, \citenamefont {Murayama},\ and\ \citenamefont
  {Pena-Garay}}]{Bahcall:2004ip}%
  \BibitemOpen
  \bibfield  {author} {\bibinfo {author} {\bibfnamefont {J.~N.}\ \bibnamefont
  {Bahcall}}, \bibinfo {author} {\bibfnamefont {H.}~\bibnamefont {Murayama}},\
  and\ \bibinfo {author} {\bibfnamefont {C.}~\bibnamefont {Pena-Garay}},\
  }\bibfield  {title} {\bibinfo {title} {{What can we learn from neutrinoless
  double beta decay experiments?}},\ }\href
  {https://doi.org/10.1103/PhysRevD.70.033012} {\bibfield  {journal} {\bibinfo
  {journal} {Phys. Rev. D}\ }\textbf {\bibinfo {volume} {70}},\ \bibinfo
  {pages} {033012} (\bibinfo {year} {2004})},\ \Eprint
  {https://arxiv.org/abs/hep-ph/0403167} {arXiv:hep-ph/0403167 [hep-ph]}
  \BibitemShut {NoStop}%
\bibitem [{\citenamefont {Wolfenstein}(1978)}]{Wolfenstein:1977ue}%
  \BibitemOpen
  \bibfield  {author} {\bibinfo {author} {\bibfnamefont {L.}~\bibnamefont
  {Wolfenstein}},\ }\bibfield  {title} {\bibinfo {title} {{Neutrino
  Oscillations in Matter}},\ }\href {https://doi.org/10.1103/PhysRevD.17.2369}
  {\bibfield  {journal} {\bibinfo  {journal} {Phys. Rev. D}\ }\textbf {\bibinfo
  {volume} {17}},\ \bibinfo {pages} {2369} (\bibinfo {year}
  {1978})}\BibitemShut {NoStop}%
\bibitem [{\citenamefont {Fukugita}\ and\ \citenamefont
  {Yanagida}(1986)}]{Fukugita:1986hr}%
  \BibitemOpen
  \bibfield  {author} {\bibinfo {author} {\bibfnamefont {M.}~\bibnamefont
  {Fukugita}}\ and\ \bibinfo {author} {\bibfnamefont {T.}~\bibnamefont
  {Yanagida}},\ }\bibfield  {title} {\bibinfo {title} {{Baryogenesis Without
  Grand Unification}},\ }\href {https://doi.org/10.1016/0370-2693(86)91126-3}
  {\bibfield  {journal} {\bibinfo  {journal} {Phys. Lett. B}\ }\textbf
  {\bibinfo {volume} {174}},\ \bibinfo {pages} {45} (\bibinfo {year}
  {1986})}\BibitemShut {NoStop}%
\bibitem [{\citenamefont {Buchmuller}\ and\ \citenamefont
  {Plumacher}(1996)}]{Buchmuller:1996pa}%
  \BibitemOpen
  \bibfield  {author} {\bibinfo {author} {\bibfnamefont {W.}~\bibnamefont
  {Buchmuller}}\ and\ \bibinfo {author} {\bibfnamefont {M.}~\bibnamefont
  {Plumacher}},\ }\bibfield  {title} {\bibinfo {title} {{Baryon asymmetry and
  neutrino mixing}},\ }\href {https://doi.org/10.1016/S0370-2693(96)01232-4}
  {\bibfield  {journal} {\bibinfo  {journal} {Phys. Lett. B}\ }\textbf
  {\bibinfo {volume} {389}},\ \bibinfo {pages} {73} (\bibinfo {year} {1996})},\
  \Eprint {https://arxiv.org/abs/hep-ph/9608308} {arXiv:hep-ph/9608308}
  \BibitemShut {NoStop}%
\bibitem [{\citenamefont {Buchmuller}\ \emph
  {et~al.}(2005{\natexlab{a}})\citenamefont {Buchmuller}, \citenamefont
  {Di~Bari},\ and\ \citenamefont {Plumacher}}]{Buchmuller:2004nz}%
  \BibitemOpen
  \bibfield  {author} {\bibinfo {author} {\bibfnamefont {W.}~\bibnamefont
  {Buchmuller}}, \bibinfo {author} {\bibfnamefont {P.}~\bibnamefont
  {Di~Bari}},\ and\ \bibinfo {author} {\bibfnamefont {M.}~\bibnamefont
  {Plumacher}},\ }\bibfield  {title} {\bibinfo {title} {{Leptogenesis for
  pedestrians}},\ }\href {https://doi.org/10.1016/j.aop.2004.02.003} {\bibfield
   {journal} {\bibinfo  {journal} {Annals Phys.}\ }\textbf {\bibinfo {volume}
  {315}},\ \bibinfo {pages} {305} (\bibinfo {year} {2005}{\natexlab{a}})},\
  \Eprint {https://arxiv.org/abs/hep-ph/0401240} {arXiv:hep-ph/0401240}
  \BibitemShut {NoStop}%
\bibitem [{\citenamefont {Buchmuller}\ \emph
  {et~al.}(2005{\natexlab{b}})\citenamefont {Buchmuller}, \citenamefont
  {Peccei},\ and\ \citenamefont {Yanagida}}]{Buchmuller:2005eh}%
  \BibitemOpen
  \bibfield  {author} {\bibinfo {author} {\bibfnamefont {W.}~\bibnamefont
  {Buchmuller}}, \bibinfo {author} {\bibfnamefont {R.~D.}\ \bibnamefont
  {Peccei}},\ and\ \bibinfo {author} {\bibfnamefont {T.}~\bibnamefont
  {Yanagida}},\ }\bibfield  {title} {\bibinfo {title} {{Leptogenesis as the
  origin of matter}},\ }\href
  {https://doi.org/10.1146/annurev.nucl.55.090704.151558} {\bibfield  {journal}
  {\bibinfo  {journal} {Ann. Rev. Nucl. Part. Sci.}\ }\textbf {\bibinfo
  {volume} {55}},\ \bibinfo {pages} {311} (\bibinfo {year}
  {2005}{\natexlab{b}})},\ \Eprint {https://arxiv.org/abs/hep-ph/0502169}
  {arXiv:hep-ph/0502169} \BibitemShut {NoStop}%
\bibitem [{\citenamefont {Pilaftsis}(1997)}]{Pilaftsis:1997jf}%
  \BibitemOpen
  \bibfield  {author} {\bibinfo {author} {\bibfnamefont {A.}~\bibnamefont
  {Pilaftsis}},\ }\bibfield  {title} {\bibinfo {title} {{CP violation and
  baryogenesis due to heavy Majorana neutrinos}},\ }\href
  {https://doi.org/10.1103/PhysRevD.56.5431} {\bibfield  {journal} {\bibinfo
  {journal} {Phys. Rev. D}\ }\textbf {\bibinfo {volume} {56}},\ \bibinfo
  {pages} {5431} (\bibinfo {year} {1997})},\ \Eprint
  {https://arxiv.org/abs/hep-ph/9707235} {arXiv:hep-ph/9707235} \BibitemShut
  {NoStop}%
\bibitem [{\citenamefont {Harrison}\ and\ \citenamefont
  {Scott}(2002)}]{Harrison:2002et}%
  \BibitemOpen
  \bibfield  {author} {\bibinfo {author} {\bibfnamefont {P.~F.}\ \bibnamefont
  {Harrison}}\ and\ \bibinfo {author} {\bibfnamefont {W.~G.}\ \bibnamefont
  {Scott}},\ }\bibfield  {title} {\bibinfo {title} {{mu - tau reflection
  symmetry in lepton mixing and neutrino oscillations}},\ }\href
  {https://doi.org/10.1016/S0370-2693(02)02772-7} {\bibfield  {journal}
  {\bibinfo  {journal} {Phys. Lett. B}\ }\textbf {\bibinfo {volume} {547}},\
  \bibinfo {pages} {219} (\bibinfo {year} {2002})},\ \Eprint
  {https://arxiv.org/abs/hep-ph/0210197} {arXiv:hep-ph/0210197} \BibitemShut
  {NoStop}%
\bibitem [{\citenamefont {Ayres}\ \emph {et~al.}(2007)\citenamefont {Ayres}
  \emph {et~al.}}]{Ayres:2007tu}%
  \BibitemOpen
  \bibfield  {author} {\bibinfo {author} {\bibfnamefont {D.~S.}\ \bibnamefont
  {Ayres}} \emph {et~al.} (\bibinfo {collaboration} {NOvA}),\ }\href
  {https://doi.org/10.2172/935497} {\bibinfo {title} {{The NOvA Technical
  Design Report}}} (\bibinfo {year} {2007})\BibitemShut {NoStop}%
\bibitem [{\citenamefont {Adamson}\ \emph {et~al.}(2016)\citenamefont {Adamson}
  \emph {et~al.}}]{Adamson:2015dkw}%
  \BibitemOpen
  \bibfield  {author} {\bibinfo {author} {\bibfnamefont {P.}~\bibnamefont
  {Adamson}} \emph {et~al.},\ }\bibfield  {title} {\bibinfo {title} {{The NuMI
  Neutrino Beam}},\ }\href {https://doi.org/10.1016/j.nima.2015.08.063}
  {\bibfield  {journal} {\bibinfo  {journal} {Nucl. Instrum. Meth. A}\ }\textbf
  {\bibinfo {volume} {806}},\ \bibinfo {pages} {279} (\bibinfo {year}
  {2016})},\ \Eprint {https://arxiv.org/abs/1507.06690} {arXiv:1507.06690
  [physics.acc-ph]} \BibitemShut {NoStop}%
\bibitem [{\citenamefont {Mufson}\ \emph {et~al.}(2015)\citenamefont {Mufson}
  \emph {et~al.}}]{Mufson:2015kga}%
  \BibitemOpen
  \bibfield  {author} {\bibinfo {author} {\bibfnamefont {S.}~\bibnamefont
  {Mufson}} \emph {et~al.},\ }\bibfield  {title} {\bibinfo {title} {{Liquid
  scintillator production for the NOvA experiment}},\ }\href
  {https://doi.org/10.1016/j.nima.2015.07.026} {\bibfield  {journal} {\bibinfo
  {journal} {Nucl. Instrum. Meth. A}\ }\textbf {\bibinfo {volume} {799}},\
  \bibinfo {pages} {1} (\bibinfo {year} {2015})},\ \Eprint
  {https://arxiv.org/abs/1504.04035} {arXiv:1504.04035 [physics.ins-det]}
  \BibitemShut {NoStop}%
\bibitem [{\citenamefont {Agostinelli}\ \emph {et~al.}(2003)\citenamefont
  {Agostinelli} \emph {et~al.}}]{Agostinelli:2002hh}%
  \BibitemOpen
  \bibfield  {author} {\bibinfo {author} {\bibfnamefont {S.}~\bibnamefont
  {Agostinelli}} \emph {et~al.} (\bibinfo {collaboration} {GEANT4}),\
  }\bibfield  {title} {\bibinfo {title} {{GEANT4--a simulation toolkit}},\
  }\href {https://doi.org/10.1016/S0168-9002(03)01368-8} {\bibfield  {journal}
  {\bibinfo  {journal} {Nucl. Instrum. Meth. A}\ }\textbf {\bibinfo {volume}
  {506}},\ \bibinfo {pages} {250} (\bibinfo {year} {2003})}\BibitemShut
  {NoStop}%
\bibitem [{\citenamefont {Aliaga}\ \emph {et~al.}(2016)\citenamefont {Aliaga}
  \emph {et~al.}}]{Aliaga:2016oaz}%
  \BibitemOpen
  \bibfield  {author} {\bibinfo {author} {\bibfnamefont {L.}~\bibnamefont
  {Aliaga}} \emph {et~al.} (\bibinfo {collaboration} {MINERvA}),\ }\bibfield
  {title} {\bibinfo {title} {{Neutrino Flux Predictions for the NuMI Beam}},\
  }\href {https://doi.org/10.1103/PhysRevD.94.092005,
  10.1103/PhysRevD.95.039903} {\bibfield  {journal} {\bibinfo  {journal} {Phys.
  Rev. D}\ }\textbf {\bibinfo {volume} {94}},\ \bibinfo {pages} {092005}
  (\bibinfo {year} {2016})},\ \bibinfo {note} {[Addendum: Phys.
  Rev.D95,no.3,039903(2017)]},\ \Eprint {https://arxiv.org/abs/1607.00704}
  {arXiv:1607.00704 [hep-ex]} \BibitemShut {NoStop}%
\bibitem [{\citenamefont {Paley}\ \emph {et~al.}(2014)\citenamefont {Paley}
  \emph {et~al.}}]{Paley:2014rpb}%
  \BibitemOpen
  \bibfield  {author} {\bibinfo {author} {\bibfnamefont {J.~M.}\ \bibnamefont
  {Paley}} \emph {et~al.} (\bibinfo {collaboration} {MIPP}),\ }\bibfield
  {title} {\bibinfo {title} {{Measurement of Charged Pion Production Yields off
  the NuMI Target}},\ }\href {https://doi.org/10.1103/PhysRevD.90.032001}
  {\bibfield  {journal} {\bibinfo  {journal} {Phys. Rev. D}\ }\textbf {\bibinfo
  {volume} {90}},\ \bibinfo {pages} {032001} (\bibinfo {year} {2014})},\
  \Eprint {https://arxiv.org/abs/1404.5882} {arXiv:1404.5882 [hep-ex]}
  \BibitemShut {NoStop}%
\bibitem [{\citenamefont {Alt}\ \emph {et~al.}(2007)\citenamefont {Alt} \emph
  {et~al.}}]{Alt:2006fr}%
  \BibitemOpen
  \bibfield  {author} {\bibinfo {author} {\bibfnamefont {C.}~\bibnamefont
  {Alt}} \emph {et~al.} (\bibinfo {collaboration} {NA49}),\ }\bibfield  {title}
  {\bibinfo {title} {{Inclusive production of charged pions in p+C collisions
  at 158-GeV/c beam momentum}},\ }\href
  {https://doi.org/10.1140/epjc/s10052-006-0165-7} {\bibfield  {journal}
  {\bibinfo  {journal} {Eur. Phys. J. C}\ }\textbf {\bibinfo {volume} {49}},\
  \bibinfo {pages} {897} (\bibinfo {year} {2007})},\ \Eprint
  {https://arxiv.org/abs/hep-ex/0606028} {arXiv:hep-ex/0606028 [hep-ex]}
  \BibitemShut {NoStop}%
\bibitem [{\citenamefont {Abgrall}\ \emph {et~al.}(2011)\citenamefont {Abgrall}
  \emph {et~al.}}]{Abgrall:2011ae}%
  \BibitemOpen
  \bibfield  {author} {\bibinfo {author} {\bibfnamefont {N.}~\bibnamefont
  {Abgrall}} \emph {et~al.} (\bibinfo {collaboration} {NA61/SHINE}),\
  }\bibfield  {title} {\bibinfo {title} {{Measurements of Cross Sections and
  Charged Pion Spectra in Proton-Carbon Interactions at 31 GeV/c}},\ }\href
  {https://doi.org/10.1103/PhysRevC.84.034604} {\bibfield  {journal} {\bibinfo
  {journal} {Phys. Rev. C}\ }\textbf {\bibinfo {volume} {84}},\ \bibinfo
  {pages} {034604} (\bibinfo {year} {2011})},\ \Eprint
  {https://arxiv.org/abs/1102.0983} {arXiv:1102.0983 [hep-ex]} \BibitemShut
  {NoStop}%
\bibitem [{\citenamefont {Barton}\ \emph {et~al.}(1983)\citenamefont {Barton}
  \emph {et~al.}}]{Barton:1982dg}%
  \BibitemOpen
  \bibfield  {author} {\bibinfo {author} {\bibfnamefont {D.~S.}\ \bibnamefont
  {Barton}} \emph {et~al.},\ }\bibfield  {title} {\bibinfo {title}
  {{Experimental Study of the a-Dependence of Inclusive Hadron
  Fragmentation}},\ }\href {https://doi.org/10.1103/PhysRevD.27.2580}
  {\bibfield  {journal} {\bibinfo  {journal} {Phys. Rev. D}\ }\textbf {\bibinfo
  {volume} {27}},\ \bibinfo {pages} {2580} (\bibinfo {year}
  {1983})}\BibitemShut {NoStop}%
\bibitem [{\citenamefont {Seun}(2007)}]{Seun:2007zz}%
  \BibitemOpen
  \bibfield  {author} {\bibinfo {author} {\bibfnamefont {S.~M.}\ \bibnamefont
  {Seun}},\ }\emph {\bibinfo {title} {{Measurement of $\pi-K$ ratios from the
  NuMI target}}},\ \href {https://doi.org/10.2172/935004} {Ph.D. thesis},\
  \bibinfo  {school} {Harvard U.} (\bibinfo {year} {2007})\BibitemShut
  {NoStop}%
\bibitem [{\citenamefont {Tinti}(2010)}]{Tinti:2010zz}%
  \BibitemOpen
  \bibfield  {author} {\bibinfo {author} {\bibfnamefont {G.~M.}\ \bibnamefont
  {Tinti}},\ }\emph {\bibinfo {title} {{Sterile neutrino oscillations in MINOS
  and hadron production in pC collisions}}},\ \href
  {https://doi.org/10.2172/992263} {Ph.D. thesis},\ \bibinfo  {school} {Oxford
  U.} (\bibinfo {year} {2010})\BibitemShut {NoStop}%
\bibitem [{\citenamefont {Lebedev}(2007)}]{Lebedev:2007zz}%
  \BibitemOpen
  \bibfield  {author} {\bibinfo {author} {\bibfnamefont {A.~V.}\ \bibnamefont
  {Lebedev}},\ }\emph {\bibinfo {title} {{Ratio of pion kaon production in
  proton carbon interactions}}},\ \href {https://doi.org/10.2172/948174} {Ph.D.
  thesis},\ \bibinfo  {school} {Harvard U.} (\bibinfo {year}
  {2007})\BibitemShut {NoStop}%
\bibitem [{\citenamefont {Baatar}\ \emph {et~al.}(2013)\citenamefont {Baatar}
  \emph {et~al.}}]{Baatar:2012fua}%
  \BibitemOpen
  \bibfield  {author} {\bibinfo {author} {\bibfnamefont {B.}~\bibnamefont
  {Baatar}} \emph {et~al.} (\bibinfo {collaboration} {NA49}),\ }\bibfield
  {title} {\bibinfo {title} {{Inclusive production of protons, anti-protons,
  neutrons, deuterons and tritons in p+C collisions at 158 GeV/c beam
  momentum}},\ }\href {https://doi.org/10.1140/epjc/s10052-013-2364-3}
  {\bibfield  {journal} {\bibinfo  {journal} {Eur. Phys. J. C}\ }\textbf
  {\bibinfo {volume} {73}},\ \bibinfo {pages} {2364} (\bibinfo {year}
  {2013})},\ \Eprint {https://arxiv.org/abs/1207.6520} {arXiv:1207.6520
  [hep-ex]} \BibitemShut {NoStop}%
\bibitem [{\citenamefont {Skubic}\ \emph {et~al.}(1978)\citenamefont {Skubic}
  \emph {et~al.}}]{Skubic:1978fi}%
  \BibitemOpen
  \bibfield  {author} {\bibinfo {author} {\bibfnamefont {P.}~\bibnamefont
  {Skubic}} \emph {et~al.},\ }\bibfield  {title} {\bibinfo {title} {{Neutral
  Strange Particle Production by 300-GeV Protons}},\ }\href
  {https://doi.org/10.1103/PhysRevD.18.3115} {\bibfield  {journal} {\bibinfo
  {journal} {Phys. Rev. D}\ }\textbf {\bibinfo {volume} {18}},\ \bibinfo
  {pages} {3115} (\bibinfo {year} {1978})}\BibitemShut {NoStop}%
\bibitem [{\citenamefont {Denisov}\ \emph {et~al.}(1973)\citenamefont
  {Denisov}, \citenamefont {Donskov}, \citenamefont {Gorin}, \citenamefont
  {Krasnokutsky}, \citenamefont {Petrukhin}, \citenamefont {Prokoshkin},\ and\
  \citenamefont {Stoyanova}}]{Denisov:1973zv}%
  \BibitemOpen
  \bibfield  {author} {\bibinfo {author} {\bibfnamefont {S.~P.}\ \bibnamefont
  {Denisov}}, \bibinfo {author} {\bibfnamefont {S.~V.}\ \bibnamefont
  {Donskov}}, \bibinfo {author} {\bibfnamefont {{\relax Yu}.~P.}\ \bibnamefont
  {Gorin}}, \bibinfo {author} {\bibfnamefont {R.~N.}\ \bibnamefont
  {Krasnokutsky}}, \bibinfo {author} {\bibfnamefont {A.~I.}\ \bibnamefont
  {Petrukhin}}, \bibinfo {author} {\bibfnamefont {{\relax Yu}.~D.}\
  \bibnamefont {Prokoshkin}},\ and\ \bibinfo {author} {\bibfnamefont {D.~A.}\
  \bibnamefont {Stoyanova}},\ }\bibfield  {title} {\bibinfo {title}
  {{Absorption cross-sections for pions, kaons, protons and anti-protons on
  complex nuclei in the 6-GeV/c to 60-GeV/c momentum range}},\ }\href
  {https://doi.org/10.1016/0550-3213(73)90351-9} {\bibfield  {journal}
  {\bibinfo  {journal} {Nucl. Phys. B}\ }\textbf {\bibinfo {volume} {61}},\
  \bibinfo {pages} {62} (\bibinfo {year} {1973})}\BibitemShut {NoStop}%
\bibitem [{\citenamefont {Carroll}\ \emph {et~al.}(1979)\citenamefont {Carroll}
  \emph {et~al.}}]{Carroll:1978hc}%
  \BibitemOpen
  \bibfield  {author} {\bibinfo {author} {\bibfnamefont {A.~S.}\ \bibnamefont
  {Carroll}} \emph {et~al.},\ }\bibfield  {title} {\bibinfo {title}
  {{Absorption Cross-Sections of $\pi^{\pm}$, $K^{\pm}$, p and $\bar{p}$ on
  Nuclei Between 60 GeV/c and 280 GeV/c}},\ }\href
  {https://doi.org/10.1016/0370-2693(79)90226-0} {\bibfield  {journal}
  {\bibinfo  {journal} {Phys. Lett. B}\ }\textbf {\bibinfo {volume} {80}},\
  \bibinfo {pages} {319} (\bibinfo {year} {1979})}\BibitemShut {NoStop}%
\bibitem [{\citenamefont {Abe}\ \emph {et~al.}(2013)\citenamefont {Abe} \emph
  {et~al.}}]{Abe:2012av}%
  \BibitemOpen
  \bibfield  {author} {\bibinfo {author} {\bibfnamefont {K.}~\bibnamefont
  {Abe}} \emph {et~al.} (\bibinfo {collaboration} {T2K}),\ }\bibfield  {title}
  {\bibinfo {title} {{T2K neutrino flux prediction}},\ }\href
  {https://doi.org/10.1103/PhysRevD.87.012001, 10.1103/PhysRevD.87.019902}
  {\bibfield  {journal} {\bibinfo  {journal} {Phys. Rev. D}\ }\textbf {\bibinfo
  {volume} {87}},\ \bibinfo {pages} {012001} (\bibinfo {year} {2013})},\
  \bibinfo {note} {[Addendum: Phys. Rev.D87,no.1,019902(2013)]},\ \Eprint
  {https://arxiv.org/abs/1211.0469} {arXiv:1211.0469 [hep-ex]} \BibitemShut
  {NoStop}%
\bibitem [{\citenamefont {Gaisser}\ \emph {et~al.}(1975)\citenamefont
  {Gaisser}, \citenamefont {Yodh}, \citenamefont {Barger},\ and\ \citenamefont
  {Halzen}}]{Gaisser:1975et}%
  \BibitemOpen
  \bibfield  {author} {\bibinfo {author} {\bibfnamefont {T.~K.}\ \bibnamefont
  {Gaisser}}, \bibinfo {author} {\bibfnamefont {G.~B.}\ \bibnamefont {Yodh}},
  \bibinfo {author} {\bibfnamefont {V.~D.}\ \bibnamefont {Barger}},\ and\
  \bibinfo {author} {\bibfnamefont {F.}~\bibnamefont {Halzen}},\ }\bibfield
  {title} {\bibinfo {title} {{On the Relation Between Proton Proton and
  Proton-Nucleus Cross-Sections at Very High-Energies}},\ }in\ \href@noop {}
  {\emph {\bibinfo {booktitle} {{Proceedings of the 14th International Cosmic
  Ray Converence}}}},\ Vol.~\bibinfo {volume} {7}\ (\bibinfo {year} {1975})\
  pp.\ \bibinfo {pages} {2161--2166}\BibitemShut {NoStop}%
\bibitem [{\citenamefont {Cronin}\ \emph {et~al.}(1957)\citenamefont {Cronin},
  \citenamefont {Cool},\ and\ \citenamefont {Abashian}}]{Cronin:1957zz}%
  \BibitemOpen
  \bibfield  {author} {\bibinfo {author} {\bibfnamefont {J.~W.}\ \bibnamefont
  {Cronin}}, \bibinfo {author} {\bibfnamefont {R.}~\bibnamefont {Cool}},\ and\
  \bibinfo {author} {\bibfnamefont {A.}~\bibnamefont {Abashian}},\ }\bibfield
  {title} {\bibinfo {title} {{Cross Sections of Nuclei for High-Energy
  Pions}},\ }\href {https://doi.org/10.1103/PhysRev.107.1121} {\bibfield
  {journal} {\bibinfo  {journal} {Phys. Rev.}\ }\textbf {\bibinfo {volume}
  {107}},\ \bibinfo {pages} {1121} (\bibinfo {year} {1957})}\BibitemShut
  {NoStop}%
\bibitem [{\citenamefont {Allaby}\ \emph {et~al.}(1969)\citenamefont {Allaby}
  \emph {et~al.}}]{Allaby:1969de}%
  \BibitemOpen
  \bibfield  {author} {\bibinfo {author} {\bibfnamefont {J.~V.}\ \bibnamefont
  {Allaby}} \emph {et~al.} (\bibinfo {collaboration} {IHEP-CERN}),\ }\bibfield
  {title} {\bibinfo {title} {{Total cross-sections of pi-minus, k-minus, and
  anti-p on protons and deuterons in the momentum range 20-65 gev/c}},\ }\href
  {https://doi.org/10.1016/0370-2693(69)90184-1} {\bibfield  {journal}
  {\bibinfo  {journal} {Phys. Lett.}\ }\textbf {\bibinfo {volume} {30B}},\
  \bibinfo {pages} {500} (\bibinfo {year} {1969})}\BibitemShut {NoStop}%
\bibitem [{\citenamefont {Longo}\ and\ \citenamefont
  {Moyer}(1962)}]{Longo:1962zz}%
  \BibitemOpen
  \bibfield  {author} {\bibinfo {author} {\bibfnamefont {M.~J.}\ \bibnamefont
  {Longo}}\ and\ \bibinfo {author} {\bibfnamefont {B.~J.}\ \bibnamefont
  {Moyer}},\ }\bibfield  {title} {\bibinfo {title} {{Nucleon and Nuclear Cross
  Sections for Positive Pions and Protons above 1.4 Bev/c}},\ }\href
  {https://doi.org/10.1103/PhysRev.125.701} {\bibfield  {journal} {\bibinfo
  {journal} {Phys. Rev.}\ }\textbf {\bibinfo {volume} {125}},\ \bibinfo {pages}
  {701} (\bibinfo {year} {1962})}\BibitemShut {NoStop}%
\bibitem [{\citenamefont {Bobchenko}\ \emph {et~al.}(1979)\citenamefont
  {Bobchenko} \emph {et~al.}}]{Bobchenko:1979hp}%
  \BibitemOpen
  \bibfield  {author} {\bibinfo {author} {\bibfnamefont {B.~M.}\ \bibnamefont
  {Bobchenko}} \emph {et~al.},\ }\bibfield  {title} {\bibinfo {title} {{
  Measurement of total inelastic cross-sections from proton interactions with
  Nuclei in the momentum range from 5 GeV/c to 9 GeV/c and pi mesons from with
  nuclei in the momentum range from 1.75 GeV/c to 6.5 GeV/c}},\ }\href@noop {}
  {\bibfield  {journal} {\bibinfo  {journal} {Sov. J. Nucl. Phys.}\ }\textbf
  {\bibinfo {volume} {30}},\ \bibinfo {pages} {805} (\bibinfo {year} {1979})},\
  \bibinfo {note} {[Yad. Fiz.30,1553(1979)]}\BibitemShut {NoStop}%
\bibitem [{\citenamefont {Fedorov}\ \emph {et~al.}(1978)\citenamefont
  {Fedorov}, \citenamefont {Grishuk}, \citenamefont {Kosov}, \citenamefont
  {Leksin}, \citenamefont {Pivnyuk}, \citenamefont {Shevchenko}, \citenamefont
  {Stolin}, \citenamefont {Vlasov},\ and\ \citenamefont
  {Vorobev}}]{Fedorov:1977an}%
  \BibitemOpen
  \bibfield  {author} {\bibinfo {author} {\bibfnamefont {V.~B.}\ \bibnamefont
  {Fedorov}}, \bibinfo {author} {\bibfnamefont {{\relax Yu}.~G.}\ \bibnamefont
  {Grishuk}}, \bibinfo {author} {\bibfnamefont {M.~V.}\ \bibnamefont {Kosov}},
  \bibinfo {author} {\bibfnamefont {G.~A.}\ \bibnamefont {Leksin}}, \bibinfo
  {author} {\bibfnamefont {N.~A.}\ \bibnamefont {Pivnyuk}}, \bibinfo {author}
  {\bibfnamefont {S.~V.}\ \bibnamefont {Shevchenko}}, \bibinfo {author}
  {\bibfnamefont {V.~L.}\ \bibnamefont {Stolin}}, \bibinfo {author}
  {\bibfnamefont {A.~V.}\ \bibnamefont {Vlasov}},\ and\ \bibinfo {author}
  {\bibfnamefont {L.~S.}\ \bibnamefont {Vorobev}},\ }\bibfield  {title}
  {\bibinfo {title} {{Total Inelastic Cross-Sections for pi Mesons on Nuclei in
  the 2-GeV/c to 6-GeV/c Momentum Range}},\ }\href@noop {} {\bibfield
  {journal} {\bibinfo  {journal} {Sov. J. Nucl. Phys.}\ }\textbf {\bibinfo
  {volume} {27}},\ \bibinfo {pages} {222} (\bibinfo {year} {1978})},\ \bibinfo
  {note} {[Yad. Fiz.27,413(1978)]}\BibitemShut {NoStop}%
\bibitem [{\citenamefont {Abrams}\ \emph {et~al.}(1970)\citenamefont {Abrams},
  \citenamefont {Cool}, \citenamefont {Giacomelli}, \citenamefont {Kycia},
  \citenamefont {Leontic}, \citenamefont {Li},\ and\ \citenamefont
  {Michael}}]{Abrams:1969jm}%
  \BibitemOpen
  \bibfield  {author} {\bibinfo {author} {\bibfnamefont {R.~J.}\ \bibnamefont
  {Abrams}}, \bibinfo {author} {\bibfnamefont {R.~L.}\ \bibnamefont {Cool}},
  \bibinfo {author} {\bibfnamefont {G.}~\bibnamefont {Giacomelli}}, \bibinfo
  {author} {\bibfnamefont {T.~F.}\ \bibnamefont {Kycia}}, \bibinfo {author}
  {\bibfnamefont {B.~A.}\ \bibnamefont {Leontic}}, \bibinfo {author}
  {\bibfnamefont {K.~K.}\ \bibnamefont {Li}},\ and\ \bibinfo {author}
  {\bibfnamefont {D.~N.}\ \bibnamefont {Michael}},\ }\bibfield  {title}
  {\bibinfo {title} {{Total cross-sections of K+- mesons and anti-protons on
  nucleons up to 3.3-GeV/c}},\ }\href {https://doi.org/10.1103/PhysRevD.1.1917}
  {\bibfield  {journal} {\bibinfo  {journal} {Phys. Rev. D}\ }\textbf {\bibinfo
  {volume} {1}},\ \bibinfo {pages} {1917} (\bibinfo {year} {1970})}\BibitemShut
  {NoStop}%
\bibitem [{\citenamefont {Andreopoulos}\ \emph {et~al.}(2010)\citenamefont
  {Andreopoulos} \emph {et~al.}}]{Andreopoulos:2009rq}%
  \BibitemOpen
  \bibfield  {author} {\bibinfo {author} {\bibfnamefont {C.}~\bibnamefont
  {Andreopoulos}} \emph {et~al.},\ }\bibfield  {title} {\bibinfo {title} {{The
  GENIE Neutrino Monte Carlo Generator}},\ }\href
  {https://doi.org/10.1016/j.nima.2009.12.009} {\bibfield  {journal} {\bibinfo
  {journal} {Nucl. Instrum. Meth. A}\ }\textbf {\bibinfo {volume} {614}},\
  \bibinfo {pages} {87} (\bibinfo {year} {2010})},\ \Eprint
  {https://arxiv.org/abs/0905.2517} {arXiv:0905.2517 [hep-ph]} \BibitemShut
  {NoStop}%
\bibitem [{\citenamefont {Andreopoulos}\ \emph {et~al.}(2015)\citenamefont
  {Andreopoulos}, \citenamefont {Barry}, \citenamefont {Dytman}, \citenamefont
  {Gallagher}, \citenamefont {Golan}, \citenamefont {Hatcher}, \citenamefont
  {Perdue},\ and\ \citenamefont {Yarba}}]{Andreopoulos:2015wxa}%
  \BibitemOpen
  \bibfield  {author} {\bibinfo {author} {\bibfnamefont {C.}~\bibnamefont
  {Andreopoulos}}, \bibinfo {author} {\bibfnamefont {C.}~\bibnamefont {Barry}},
  \bibinfo {author} {\bibfnamefont {S.}~\bibnamefont {Dytman}}, \bibinfo
  {author} {\bibfnamefont {H.}~\bibnamefont {Gallagher}}, \bibinfo {author}
  {\bibfnamefont {T.}~\bibnamefont {Golan}}, \bibinfo {author} {\bibfnamefont
  {R.}~\bibnamefont {Hatcher}}, \bibinfo {author} {\bibfnamefont
  {G.}~\bibnamefont {Perdue}},\ and\ \bibinfo {author} {\bibfnamefont
  {J.}~\bibnamefont {Yarba}},\ }\href@noop {} {\bibinfo {title} {{The GENIE
  Neutrino Monte Carlo Generator: Physics and User Manual}}} (\bibinfo {year}
  {2015}),\ \Eprint {https://arxiv.org/abs/1510.05494} {arXiv:1510.05494
  [hep-ph]} \BibitemShut {NoStop}%
\bibitem [{\citenamefont {Tena-Vidal}\ \emph {et~al.}(2021)\citenamefont
  {Tena-Vidal} \emph {et~al.}}]{GENIE:2021zuu}%
  \BibitemOpen
  \bibfield  {author} {\bibinfo {author} {\bibfnamefont {J.}~\bibnamefont
  {Tena-Vidal}} \emph {et~al.} (\bibinfo {collaboration} {GENIE}),\ }\bibfield
  {title} {\bibinfo {title} {{Neutrino-Nucleon Cross-Section Model Tuning in
  GENIE v3}},\ }\Eprint {https://arxiv.org/abs/2104.09179} {arXiv:2104.09179
  [hep-ph]}  (\bibinfo {year} {2021}),\ \bibinfo {note}
  {{FERMILAB-PUB-20-531-SCD-T}}\BibitemShut {NoStop}%
\bibitem [{\citenamefont {Tena-Vidal}(2018)}]{Tena-Vidal:2018misc}%
  \BibitemOpen
  \bibfield  {author} {\bibinfo {author} {\bibfnamefont {J.}~\bibnamefont
  {Tena-Vidal}},\ }\href
  {https://indico.cern.ch/event/703880/contributions/3157410/attachments/1734479/2804781/Tena_vidal.pdf}
  {\bibinfo {title} {{Tuning the pion production with GENIE version 3}}},\
  \bibinfo {howpublished} {{NuInt18, 12th International Workshop on
  Neutrino-Nucleus Interactions in the Few GeV Region}} (\bibinfo {year}
  {2018})\BibitemShut {NoStop}%
\bibitem [{\citenamefont {Nieves}\ \emph {et~al.}(2004)\citenamefont {Nieves},
  \citenamefont {Amaro},\ and\ \citenamefont {Valverde}}]{Nieves:2004wx}%
  \BibitemOpen
  \bibfield  {author} {\bibinfo {author} {\bibfnamefont {J.}~\bibnamefont
  {Nieves}}, \bibinfo {author} {\bibfnamefont {J.~E.}\ \bibnamefont {Amaro}},\
  and\ \bibinfo {author} {\bibfnamefont {M.}~\bibnamefont {Valverde}},\
  }\bibfield  {title} {\bibinfo {title} {{Inclusive quasi-elastic neutrino
  reactions}},\ }\href {https://doi.org/10.1103/PhysRevC.70.055503,
  10.1103/PhysRevC.72.019902} {\bibfield  {journal} {\bibinfo  {journal} {Phys.
  Rev. C}\ }\textbf {\bibinfo {volume} {70}},\ \bibinfo {pages} {055503}
  (\bibinfo {year} {2004})},\ \bibinfo {note} {[Erratum: Phys.
  Rev.C72,019902(2005)]},\ \Eprint {https://arxiv.org/abs/nucl-th/0408005}
  {arXiv:nucl-th/0408005 [nucl-th]} \BibitemShut {NoStop}%
\bibitem [{\citenamefont {Martini}\ \emph {et~al.}(2009)\citenamefont
  {Martini}, \citenamefont {Ericson}, \citenamefont {Chanfray},\ and\
  \citenamefont {Marteau}}]{Martini:2009uj}%
  \BibitemOpen
  \bibfield  {author} {\bibinfo {author} {\bibfnamefont {M.}~\bibnamefont
  {Martini}}, \bibinfo {author} {\bibfnamefont {M.}~\bibnamefont {Ericson}},
  \bibinfo {author} {\bibfnamefont {G.}~\bibnamefont {Chanfray}},\ and\
  \bibinfo {author} {\bibfnamefont {J.}~\bibnamefont {Marteau}},\ }\bibfield
  {title} {\bibinfo {title} {{A Unified approach for nucleon knock-out,
  coherent and incoherent pion production in neutrino interactions with
  nuclei}},\ }\href {https://doi.org/10.1103/PhysRevC.80.065501} {\bibfield
  {journal} {\bibinfo  {journal} {Phys. Rev. C}\ }\textbf {\bibinfo {volume}
  {80}},\ \bibinfo {pages} {065501} (\bibinfo {year} {2009})},\ \Eprint
  {https://arxiv.org/abs/0910.2622} {arXiv:0910.2622 [nucl-th]} \BibitemShut
  {NoStop}%
\bibitem [{\citenamefont {Pandey}\ \emph {et~al.}(2015)\citenamefont {Pandey},
  \citenamefont {Jachowicz}, \citenamefont {Van~Cuyck}, \citenamefont
  {Ryckebusch},\ and\ \citenamefont {Martini}}]{Pandey:2014tza}%
  \BibitemOpen
  \bibfield  {author} {\bibinfo {author} {\bibfnamefont {V.}~\bibnamefont
  {Pandey}}, \bibinfo {author} {\bibfnamefont {N.}~\bibnamefont {Jachowicz}},
  \bibinfo {author} {\bibfnamefont {T.}~\bibnamefont {Van~Cuyck}}, \bibinfo
  {author} {\bibfnamefont {J.}~\bibnamefont {Ryckebusch}},\ and\ \bibinfo
  {author} {\bibfnamefont {M.}~\bibnamefont {Martini}},\ }\bibfield  {title}
  {\bibinfo {title} {{Low-energy excitations and quasielastic contribution to
  electron-nucleus and neutrino-nucleus scattering in the continuum
  random-phase approximation}},\ }\href
  {https://doi.org/10.1103/PhysRevC.92.024606} {\bibfield  {journal} {\bibinfo
  {journal} {Phys. Rev. C}\ }\textbf {\bibinfo {volume} {92}},\ \bibinfo
  {pages} {024606} (\bibinfo {year} {2015})},\ \Eprint
  {https://arxiv.org/abs/1412.4624} {arXiv:1412.4624 [nucl-th]} \BibitemShut
  {NoStop}%
\bibitem [{\citenamefont {Meyer}\ \emph {et~al.}(2016)\citenamefont {Meyer},
  \citenamefont {Betancourt}, \citenamefont {Gran},\ and\ \citenamefont
  {Hill}}]{Meyer:2016oeg}%
  \BibitemOpen
  \bibfield  {author} {\bibinfo {author} {\bibfnamefont {A.~S.}\ \bibnamefont
  {Meyer}}, \bibinfo {author} {\bibfnamefont {M.}~\bibnamefont {Betancourt}},
  \bibinfo {author} {\bibfnamefont {R.}~\bibnamefont {Gran}},\ and\ \bibinfo
  {author} {\bibfnamefont {R.~J.}\ \bibnamefont {Hill}},\ }\bibfield  {title}
  {\bibinfo {title} {{Deuterium target data for precision neutrino-nucleus
  cross sections}},\ }\href {https://doi.org/10.1103/PhysRevD.93.113015}
  {\bibfield  {journal} {\bibinfo  {journal} {Phys. Rev. D}\ }\textbf {\bibinfo
  {volume} {93}},\ \bibinfo {pages} {113015} (\bibinfo {year} {2016})},\
  \Eprint {https://arxiv.org/abs/1603.03048} {arXiv:1603.03048 [hep-ph]}
  \BibitemShut {NoStop}%
\bibitem [{\citenamefont {Nieves}\ \emph {et~al.}(2011)\citenamefont {Nieves},
  \citenamefont {Ruiz~Simo},\ and\ \citenamefont
  {Vicente~Vacas}}]{Nieves:2011pp}%
  \BibitemOpen
  \bibfield  {author} {\bibinfo {author} {\bibfnamefont {J.}~\bibnamefont
  {Nieves}}, \bibinfo {author} {\bibfnamefont {I.}~\bibnamefont {Ruiz~Simo}},\
  and\ \bibinfo {author} {\bibfnamefont {M.~J.}\ \bibnamefont
  {Vicente~Vacas}},\ }\bibfield  {title} {\bibinfo {title} {{Inclusive
  Charged--Current Neutrino--Nucleus Reactions}},\ }\href
  {https://doi.org/10.1103/PhysRevC.83.045501} {\bibfield  {journal} {\bibinfo
  {journal} {Phys. Rev. C}\ }\textbf {\bibinfo {volume} {83}},\ \bibinfo
  {pages} {045501} (\bibinfo {year} {2011})},\ \Eprint
  {https://arxiv.org/abs/1102.2777} {arXiv:1102.2777 [hep-ph]} \BibitemShut
  {NoStop}%
\bibitem [{\citenamefont {Gran}\ \emph {et~al.}(2013)\citenamefont {Gran},
  \citenamefont {Nieves}, \citenamefont {Sanchez},\ and\ \citenamefont
  {Vicente~Vacas}}]{Gran:2013kda}%
  \BibitemOpen
  \bibfield  {author} {\bibinfo {author} {\bibfnamefont {R.}~\bibnamefont
  {Gran}}, \bibinfo {author} {\bibfnamefont {J.}~\bibnamefont {Nieves}},
  \bibinfo {author} {\bibfnamefont {F.}~\bibnamefont {Sanchez}},\ and\ \bibinfo
  {author} {\bibfnamefont {M.~J.}\ \bibnamefont {Vicente~Vacas}},\ }\bibfield
  {title} {\bibinfo {title} {{Neutrino-nucleus quasi-elastic and 2p2h
  interactions up to 10 GeV}},\ }\href
  {https://doi.org/10.1103/PhysRevD.88.113007} {\bibfield  {journal} {\bibinfo
  {journal} {Phys. Rev. D}\ }\textbf {\bibinfo {volume} {88}},\ \bibinfo
  {pages} {113007} (\bibinfo {year} {2013})},\ \Eprint
  {https://arxiv.org/abs/1307.8105} {arXiv:1307.8105 [hep-ph]} \BibitemShut
  {NoStop}%
\bibitem [{\citenamefont {Berger}\ and\ \citenamefont
  {Sehgal}(2007)}]{Berger:2007rq}%
  \BibitemOpen
  \bibfield  {author} {\bibinfo {author} {\bibfnamefont {C.}~\bibnamefont
  {Berger}}\ and\ \bibinfo {author} {\bibfnamefont {L.~M.}\ \bibnamefont
  {Sehgal}},\ }\bibfield  {title} {\bibinfo {title} {{Lepton mass effects in
  single pion production by neutrinos}},\ }\href
  {https://doi.org/10.1103/PhysRevD.76.113004} {\bibfield  {journal} {\bibinfo
  {journal} {Phys. Rev. D}\ }\textbf {\bibinfo {volume} {76}},\ \bibinfo
  {pages} {113004} (\bibinfo {year} {2007})},\ \Eprint
  {https://arxiv.org/abs/0709.4378} {arXiv:0709.4378 [hep-ph]} \BibitemShut
  {NoStop}%
\bibitem [{\citenamefont {Berger}\ and\ \citenamefont
  {Sehgal}(2009)}]{Berger:2008xs}%
  \BibitemOpen
  \bibfield  {author} {\bibinfo {author} {\bibfnamefont {C.}~\bibnamefont
  {Berger}}\ and\ \bibinfo {author} {\bibfnamefont {L.~M.}\ \bibnamefont
  {Sehgal}},\ }\bibfield  {title} {\bibinfo {title} {{PCAC and coherent pion
  production by low energy neutrinos}},\ }\href
  {https://doi.org/10.1103/PhysRevD.79.053003} {\bibfield  {journal} {\bibinfo
  {journal} {Phys. Rev. D}\ }\textbf {\bibinfo {volume} {79}},\ \bibinfo
  {pages} {053003} (\bibinfo {year} {2009})},\ \Eprint
  {https://arxiv.org/abs/0812.2653} {arXiv:0812.2653 [hep-ph]} \BibitemShut
  {NoStop}%
\bibitem [{\citenamefont {Bodek}\ and\ \citenamefont
  {Yang}(2003)}]{Bodek:2002ps}%
  \BibitemOpen
  \bibfield  {author} {\bibinfo {author} {\bibfnamefont {A.}~\bibnamefont
  {Bodek}}\ and\ \bibinfo {author} {\bibfnamefont {U.~K.}\ \bibnamefont
  {Yang}},\ }\bibfield  {title} {\bibinfo {title} {{Higher twist, xi(omega)
  scaling, and effective LO PDFs for lepton scattering in the few GeV
  region}},\ }\bibfield  {booktitle} {\emph {\bibinfo {booktitle} {{Neutrino
  factories. Proceedings, 4th International Workshop, NuFact'02, London, UK,
  July 1-6, 2002}}},\ }\href {https://doi.org/10.1088/0954-3899/29/8/369}
  {\bibfield  {journal} {\bibinfo  {journal} {J. Phys. G}\ }\textbf {\bibinfo
  {volume} {29}},\ \bibinfo {pages} {1899} (\bibinfo {year} {2003})},\ \Eprint
  {https://arxiv.org/abs/hep-ex/0210024} {arXiv:hep-ex/0210024 [hep-ex]}
  \BibitemShut {NoStop}%
\bibitem [{\citenamefont {Yang}\ \emph {et~al.}(2009)\citenamefont {Yang},
  \citenamefont {Andreopoulos}, \citenamefont {Gallagher}, \citenamefont
  {Hoffmann},\ and\ \citenamefont {Kehayias}}]{Yang:2009zx}%
  \BibitemOpen
  \bibfield  {author} {\bibinfo {author} {\bibfnamefont {T.}~\bibnamefont
  {Yang}}, \bibinfo {author} {\bibfnamefont {C.}~\bibnamefont {Andreopoulos}},
  \bibinfo {author} {\bibfnamefont {H.}~\bibnamefont {Gallagher}}, \bibinfo
  {author} {\bibfnamefont {K.}~\bibnamefont {Hoffmann}},\ and\ \bibinfo
  {author} {\bibfnamefont {P.}~\bibnamefont {Kehayias}},\ }\bibfield  {title}
  {\bibinfo {title} {{A Hadronization Model for Few-GeV Neutrino
  Interactions}},\ }\href {https://doi.org/10.1140/epjc/s10052-009-1094-z}
  {\bibfield  {journal} {\bibinfo  {journal} {Eur. Phys. J. C}\ }\textbf
  {\bibinfo {volume} {63}},\ \bibinfo {pages} {1} (\bibinfo {year} {2009})},\
  \Eprint {https://arxiv.org/abs/0904.4043} {arXiv:0904.4043 [hep-ph]}
  \BibitemShut {NoStop}%
\bibitem [{\citenamefont {Sjostrand}\ \emph {et~al.}(2006)\citenamefont
  {Sjostrand}, \citenamefont {Mrenna},\ and\ \citenamefont
  {Skands}}]{Sjostrand:2006za}%
  \BibitemOpen
  \bibfield  {author} {\bibinfo {author} {\bibfnamefont {T.}~\bibnamefont
  {Sjostrand}}, \bibinfo {author} {\bibfnamefont {S.}~\bibnamefont {Mrenna}},\
  and\ \bibinfo {author} {\bibfnamefont {P.~Z.}\ \bibnamefont {Skands}},\
  }\bibfield  {title} {\bibinfo {title} {{PYTHIA 6.4 Physics and Manual}},\
  }\href {https://doi.org/10.1088/1126-6708/2006/05/026} {\bibfield  {journal}
  {\bibinfo  {journal} {JHEP}\ }\textbf {\bibinfo {volume} {05}},\ \bibinfo
  {pages} {26}},\ \Eprint {https://arxiv.org/abs/hep-ph/0603175}
  {arXiv:hep-ph/0603175 [hep-ph]} \BibitemShut {NoStop}%
\bibitem [{\citenamefont {Salcedo}\ \emph {et~al.}(1988)\citenamefont
  {Salcedo}, \citenamefont {Oset}, \citenamefont {Vicente-Vacas},\ and\
  \citenamefont {Garcia-Recio}}]{Salcedo:1987md}%
  \BibitemOpen
  \bibfield  {author} {\bibinfo {author} {\bibfnamefont {L.~L.}\ \bibnamefont
  {Salcedo}}, \bibinfo {author} {\bibfnamefont {E.}~\bibnamefont {Oset}},
  \bibinfo {author} {\bibfnamefont {M.~J.}\ \bibnamefont {Vicente-Vacas}},\
  and\ \bibinfo {author} {\bibfnamefont {C.}~\bibnamefont {Garcia-Recio}},\
  }\bibfield  {title} {\bibinfo {title} {{Computer Simulation of Inclusive Pion
  Nuclear Reactions}},\ }\href {https://doi.org/10.1016/0375-9474(88)90310-7}
  {\bibfield  {journal} {\bibinfo  {journal} {Nucl. Phys. A}\ }\textbf
  {\bibinfo {volume} {484}},\ \bibinfo {pages} {557} (\bibinfo {year}
  {1988})}\BibitemShut {NoStop}%
\bibitem [{\citenamefont {Gran}\ \emph {et~al.}(2018)\citenamefont {Gran} \emph
  {et~al.}}]{Gran:2018fxa}%
  \BibitemOpen
  \bibfield  {author} {\bibinfo {author} {\bibfnamefont {R.}~\bibnamefont
  {Gran}} \emph {et~al.} (\bibinfo {collaboration} {MINERvA}),\ }\bibfield
  {title} {\bibinfo {title} {{Antineutrino Charged-Current Reactions on
  Hydrocarbon with Low Momentum Transfer}},\ }\href
  {https://doi.org/10.1103/PhysRevLett.120.221805} {\bibfield  {journal}
  {\bibinfo  {journal} {Phys. Rev. Lett.}\ }\textbf {\bibinfo {volume} {120}},\
  \bibinfo {pages} {221805} (\bibinfo {year} {2018})},\ \Eprint
  {https://arxiv.org/abs/1803.09377} {arXiv:1803.09377 [hep-ex]} \BibitemShut
  {NoStop}%
\bibitem [{\citenamefont {Allardyce}\ \emph {et~al.}(1973)\citenamefont
  {Allardyce} \emph {et~al.}}]{Allardyce:1973ce}%
  \BibitemOpen
  \bibfield  {author} {\bibinfo {author} {\bibfnamefont {B.~W.}\ \bibnamefont
  {Allardyce}} \emph {et~al.},\ }\bibfield  {title} {\bibinfo {title} {{Pion
  reaction cross-sections and nuclear sizes}},\ }\href
  {https://doi.org/10.1016/0375-9474(73)90049-3} {\bibfield  {journal}
  {\bibinfo  {journal} {Nucl. Phys. A}\ }\textbf {\bibinfo {volume} {209}},\
  \bibinfo {pages} {1} (\bibinfo {year} {1973})}\BibitemShut {NoStop}%
\bibitem [{\citenamefont {Saunders}\ \emph {et~al.}(1996)\citenamefont
  {Saunders}, \citenamefont {Hoeibraten}, \citenamefont {Kraushaar},
  \citenamefont {Kriss}, \citenamefont {Peterson}, \citenamefont {Ristinen},
  \citenamefont {Brack}, \citenamefont {Hofman}, \citenamefont {Gibson},\ and\
  \citenamefont {Morris}}]{Saunders:1996ic}%
  \BibitemOpen
  \bibfield  {author} {\bibinfo {author} {\bibfnamefont {A.}~\bibnamefont
  {Saunders}}, \bibinfo {author} {\bibfnamefont {S.}~\bibnamefont
  {Hoeibraten}}, \bibinfo {author} {\bibfnamefont {J.~J.}\ \bibnamefont
  {Kraushaar}}, \bibinfo {author} {\bibfnamefont {B.~J.}\ \bibnamefont
  {Kriss}}, \bibinfo {author} {\bibfnamefont {R.~J.}\ \bibnamefont {Peterson}},
  \bibinfo {author} {\bibfnamefont {R.~A.}\ \bibnamefont {Ristinen}}, \bibinfo
  {author} {\bibfnamefont {J.~T.}\ \bibnamefont {Brack}}, \bibinfo {author}
  {\bibfnamefont {G.}~\bibnamefont {Hofman}}, \bibinfo {author} {\bibfnamefont
  {E.~F.}\ \bibnamefont {Gibson}},\ and\ \bibinfo {author} {\bibfnamefont
  {C.~L.}\ \bibnamefont {Morris}},\ }\bibfield  {title} {\bibinfo {title}
  {{Reaction and total cross-sections for low-energy pi+ and pi- on isospin
  zero nuclei}},\ }\href {https://doi.org/10.1103/PhysRevC.53.1745} {\bibfield
  {journal} {\bibinfo  {journal} {Phys. Rev. C}\ }\textbf {\bibinfo {volume}
  {53}},\ \bibinfo {pages} {1745} (\bibinfo {year} {1996})}\BibitemShut
  {NoStop}%
\bibitem [{\citenamefont {Meirav}\ \emph {et~al.}(1989)\citenamefont {Meirav},
  \citenamefont {Friedman}, \citenamefont {Johnson}, \citenamefont
  {Olszewski},\ and\ \citenamefont {Weber}}]{Meirav:1988pn}%
  \BibitemOpen
  \bibfield  {author} {\bibinfo {author} {\bibfnamefont {O.}~\bibnamefont
  {Meirav}}, \bibinfo {author} {\bibfnamefont {E.}~\bibnamefont {Friedman}},
  \bibinfo {author} {\bibfnamefont {R.~R.}\ \bibnamefont {Johnson}}, \bibinfo
  {author} {\bibfnamefont {R.}~\bibnamefont {Olszewski}},\ and\ \bibinfo
  {author} {\bibfnamefont {P.}~\bibnamefont {Weber}},\ }\bibfield  {title}
  {\bibinfo {title} {{Low-energy Pion - Nucleus Potentials From Differential
  and Integral Data}},\ }\href {https://doi.org/10.1103/PhysRevC.40.843}
  {\bibfield  {journal} {\bibinfo  {journal} {Phys. Rev. C}\ }\textbf {\bibinfo
  {volume} {40}},\ \bibinfo {pages} {843} (\bibinfo {year} {1989})}\BibitemShut
  {NoStop}%
\bibitem [{\citenamefont {Levenson}\ \emph {et~al.}(1983)\citenamefont
  {Levenson} \emph {et~al.}}]{Levenson:1983xu}%
  \BibitemOpen
  \bibfield  {author} {\bibinfo {author} {\bibfnamefont {S.~M.}\ \bibnamefont
  {Levenson}} \emph {et~al.},\ }\bibfield  {title} {\bibinfo {title}
  {{Inclusive pion scattering in the delta (1232) region}},\ }\href
  {https://doi.org/10.1103/PhysRevC.28.326} {\bibfield  {journal} {\bibinfo
  {journal} {Phys. Rev. C}\ }\textbf {\bibinfo {volume} {28}},\ \bibinfo
  {pages} {326} (\bibinfo {year} {1983})}\BibitemShut {NoStop}%
\bibitem [{\citenamefont {Ashery}\ \emph {et~al.}(1981)\citenamefont {Ashery},
  \citenamefont {Navon}, \citenamefont {Azuelos}, \citenamefont {Walter},
  \citenamefont {Pfeiffer},\ and\ \citenamefont {Schleputz}}]{Ashery:1981tq}%
  \BibitemOpen
  \bibfield  {author} {\bibinfo {author} {\bibfnamefont {D.}~\bibnamefont
  {Ashery}}, \bibinfo {author} {\bibfnamefont {I.}~\bibnamefont {Navon}},
  \bibinfo {author} {\bibfnamefont {G.}~\bibnamefont {Azuelos}}, \bibinfo
  {author} {\bibfnamefont {H.~K.}\ \bibnamefont {Walter}}, \bibinfo {author}
  {\bibfnamefont {H.~J.}\ \bibnamefont {Pfeiffer}},\ and\ \bibinfo {author}
  {\bibfnamefont {F.~W.}\ \bibnamefont {Schleputz}},\ }\bibfield  {title}
  {\bibinfo {title} {{True Absorption and Scattering of Pions on Nuclei}},\
  }\href {https://doi.org/10.1103/PhysRevC.23.2173} {\bibfield  {journal}
  {\bibinfo  {journal} {Phys. Rev. C}\ }\textbf {\bibinfo {volume} {23}},\
  \bibinfo {pages} {2173} (\bibinfo {year} {1981})}\BibitemShut {NoStop}%
\bibitem [{\citenamefont {Ashery}\ \emph {et~al.}(1984)\citenamefont {Ashery}
  \emph {et~al.}}]{Ashery:1984ne}%
  \BibitemOpen
  \bibfield  {author} {\bibinfo {author} {\bibfnamefont {D.}~\bibnamefont
  {Ashery}} \emph {et~al.},\ }\bibfield  {title} {\bibinfo {title} {{Inclusive
  pion single charge exchange reactions}},\ }\href
  {https://doi.org/10.1103/PhysRevC.30.946} {\bibfield  {journal} {\bibinfo
  {journal} {Phys. Rev. C}\ }\textbf {\bibinfo {volume} {30}},\ \bibinfo
  {pages} {946} (\bibinfo {year} {1984})}\BibitemShut {NoStop}%
\bibitem [{\citenamefont {Pinzon~Guerra}\ \emph {et~al.}(2017)\citenamefont
  {Pinzon~Guerra} \emph {et~al.}}]{PinzonGuerra:2016uae}%
  \BibitemOpen
  \bibfield  {author} {\bibinfo {author} {\bibfnamefont {E.~S.}\ \bibnamefont
  {Pinzon~Guerra}} \emph {et~al.} (\bibinfo {collaboration} {DUET}),\
  }\bibfield  {title} {\bibinfo {title} {{Measurement of
  $\sigma_{\mathrm{ABS}}$ and $\sigma_{\mathrm{CX}}$ of $\pi^+$ on carbon by
  the Dual Use Experiment at TRIUMF (DUET)}},\ }\href
  {https://doi.org/10.1103/PhysRevC.95.045203} {\bibfield  {journal} {\bibinfo
  {journal} {Phys. Rev. C}\ }\textbf {\bibinfo {volume} {95}},\ \bibinfo
  {pages} {045203} (\bibinfo {year} {2017})},\ \Eprint
  {https://arxiv.org/abs/1611.05612} {arXiv:1611.05612 [hep-ex]} \BibitemShut
  {NoStop}%
\bibitem [{\citenamefont {{Geant4 Collaboration}}(2017)}]{Geant:2017ats}%
  \BibitemOpen
  \bibfield  {author} {\bibinfo {author} {\bibnamefont {{Geant4
  Collaboration}}},\ }\bibfield  {title} {\bibinfo {title} {Geant4 10.4 release
  notes},\ }\href
  {https://geant4-data.web.cern.ch/ReleaseNotes/ReleaseNotes4.10.4.html}
  {\bibfield  {journal} {\bibinfo  {journal} {geant4-data.web.cern.ch,
  https://geant4-data.web.cern.ch/ ReleaseNotes/ReleaseNotes4.10.4.html}\ }
  (\bibinfo {year} {2017})}\BibitemShut {NoStop}%
\bibitem [{\citenamefont {Aurisano}\ \emph {et~al.}(2015)\citenamefont
  {Aurisano}, \citenamefont {Backhouse}, \citenamefont {Hatcher}, \citenamefont
  {Mayer}, \citenamefont {Musser}, \citenamefont {Patterson}, \citenamefont
  {Schroeter},\ and\ \citenamefont {Sousa}}]{Aurisano:2015oxj}%
  \BibitemOpen
  \bibfield  {author} {\bibinfo {author} {\bibfnamefont {A.}~\bibnamefont
  {Aurisano}}, \bibinfo {author} {\bibfnamefont {C.}~\bibnamefont {Backhouse}},
  \bibinfo {author} {\bibfnamefont {R.}~\bibnamefont {Hatcher}}, \bibinfo
  {author} {\bibfnamefont {N.}~\bibnamefont {Mayer}}, \bibinfo {author}
  {\bibfnamefont {J.}~\bibnamefont {Musser}}, \bibinfo {author} {\bibfnamefont
  {R.}~\bibnamefont {Patterson}}, \bibinfo {author} {\bibfnamefont
  {R.}~\bibnamefont {Schroeter}},\ and\ \bibinfo {author} {\bibfnamefont
  {A.}~\bibnamefont {Sousa}} (\bibinfo {collaboration} {NOvA}),\ }\bibfield
  {title} {\bibinfo {title} {{The NOvA simulation chain}},\ }\bibfield
  {booktitle} {\emph {\bibinfo {booktitle} {{Proceedings, 21st International
  Conference on Computing in High Energy and Nuclear Physics (CHEP 2015):
  Okinawa, Japan, April 13-17, 2015}}},\ }\href
  {https://doi.org/10.1088/1742-6596/664/7/072002} {\bibfield  {journal}
  {\bibinfo  {journal} {J. Phys. Conf. Ser.}\ }\textbf {\bibinfo {volume}
  {664}},\ \bibinfo {pages} {072002} (\bibinfo {year} {2015})}\BibitemShut
  {NoStop}%
\bibitem [{\citenamefont {Sternheimer}(1952)}]{Sternheimer:1952jn}%
  \BibitemOpen
  \bibfield  {author} {\bibinfo {author} {\bibfnamefont {R.~M.}\ \bibnamefont
  {Sternheimer}},\ }\bibfield  {title} {\bibinfo {title} {{The Density Effect
  for Ionization Loss in Materials}},\ }\href
  {https://doi.org/10.1103/PhysRev.88.851} {\bibfield  {journal} {\bibinfo
  {journal} {Phys. Rev.}\ }\textbf {\bibinfo {volume} {88}},\ \bibinfo {pages}
  {851} (\bibinfo {year} {1952})}\BibitemShut {NoStop}%
\bibitem [{\citenamefont {Singh}(2019)}]{Singh:2019lqg}%
  \BibitemOpen
  \bibfield  {author} {\bibinfo {author} {\bibfnamefont {P.}~\bibnamefont
  {Singh}},\ }\emph {\bibinfo {title} {{Extraction of Neutrino Oscillation
  Parameters using a Simultaneous Fit of $\nu_{\mu}$ Disappearance and
  $\nu_{e}$ Appearance data with the NOvA Experiment}}},\ \href
  {https://doi.org/10.2172/1764076} {Ph.D. thesis},\ \bibinfo  {school} {Delhi
  U.} (\bibinfo {year} {2019})\BibitemShut {NoStop}%
\bibitem [{\citenamefont {Pershey}(2018)}]{Pershey:2018gtf}%
  \BibitemOpen
  \bibfield  {author} {\bibinfo {author} {\bibfnamefont {D.~S.}\ \bibnamefont
  {Pershey}},\ }\emph {\bibinfo {title} {{A Measurement of $\nu_e$ Appearance
  and $\nu_\mu$ Disappearance Neutrino Oscillations with the NOvA
  Experiment}}},\ \href {https://doi.org/10.2172/1484186} {Ph.D. thesis},\
  \bibinfo  {school} {Caltech} (\bibinfo {year} {2018})\BibitemShut {NoStop}%
\bibitem [{\citenamefont {Aurisano}\ \emph {et~al.}(2016)\citenamefont
  {Aurisano}, \citenamefont {Radovic}, \citenamefont {Rocco}, \citenamefont
  {Himmel}, \citenamefont {Messier}, \citenamefont {Niner}, \citenamefont
  {Pawloski}, \citenamefont {Psihas}, \citenamefont {Sousa},\ and\
  \citenamefont {Vahle}}]{Aurisano:2016jvx}%
  \BibitemOpen
  \bibfield  {author} {\bibinfo {author} {\bibfnamefont {A.}~\bibnamefont
  {Aurisano}}, \bibinfo {author} {\bibfnamefont {A.}~\bibnamefont {Radovic}},
  \bibinfo {author} {\bibfnamefont {D.}~\bibnamefont {Rocco}}, \bibinfo
  {author} {\bibfnamefont {A.}~\bibnamefont {Himmel}}, \bibinfo {author}
  {\bibfnamefont {M.~D.}\ \bibnamefont {Messier}}, \bibinfo {author}
  {\bibfnamefont {E.}~\bibnamefont {Niner}}, \bibinfo {author} {\bibfnamefont
  {G.}~\bibnamefont {Pawloski}}, \bibinfo {author} {\bibfnamefont
  {F.}~\bibnamefont {Psihas}}, \bibinfo {author} {\bibfnamefont
  {A.}~\bibnamefont {Sousa}},\ and\ \bibinfo {author} {\bibfnamefont
  {P.}~\bibnamefont {Vahle}},\ }\bibfield  {title} {\bibinfo {title} {{A
  Convolutional Neural Network Neutrino Event Classifier}},\ }\href
  {https://doi.org/10.1088/1748-0221/11/09/P09001} {\bibfield  {journal}
  {\bibinfo  {journal} {JINST}\ }\textbf {\bibinfo {volume} {11}}\bibfield
  {number} {\bibinfo  {number} { (09)},\ \bibinfo {pages} {P09001}},\ }\Eprint
  {https://arxiv.org/abs/1604.01444} {arXiv:1604.01444 [hep-ex]} \BibitemShut
  {NoStop}%
\bibitem [{\citenamefont {Groh}(2021)}]{Groh:2021qyu}%
  \BibitemOpen
  \bibfield  {author} {\bibinfo {author} {\bibfnamefont {M.}~\bibnamefont
  {Groh}},\ }\emph {\bibinfo {title} {{Constraints on Neutrino Oscillation
  Parameters from Neutrinos and Antineutrinos with Machine Learning}}},\
  \href@noop {} {Ph.D. thesis},\ \bibinfo  {school} {Indiana U., Bloomington
  (main)} (\bibinfo {year} {2021})\BibitemShut {NoStop}%
\bibitem [{\citenamefont {Psihas}(2018)}]{Psihas:2018czu}%
  \BibitemOpen
  \bibfield  {author} {\bibinfo {author} {\bibfnamefont {F.}~\bibnamefont
  {Psihas}},\ }\emph {\bibinfo {title} {{Measurement of Long Baseline Neutrino
  Oscillations and Improvements from Deep Learning}}},\ \href
  {https://doi.org/10.2172/1437288} {Ph.D. thesis},\ \bibinfo  {school}
  {Indiana U.} (\bibinfo {year} {2018})\BibitemShut {NoStop}%
\bibitem [{\citenamefont {Bassin}\ \emph {et~al.}(2000)\citenamefont {Bassin},
  \citenamefont {Laske},\ and\ \citenamefont {Masters}}]{Bassin:2000ats}%
  \BibitemOpen
  \bibfield  {author} {\bibinfo {author} {\bibfnamefont {C.}~\bibnamefont
  {Bassin}}, \bibinfo {author} {\bibfnamefont {G.}~\bibnamefont {Laske}},\ and\
  \bibinfo {author} {\bibfnamefont {G.}~\bibnamefont {Masters}},\ }\bibfield
  {title} {\bibinfo {title} {The current limits of resolution for surface wave
  tomography in north america},\ }\href@noop {} {\bibfield  {journal} {\bibinfo
   {journal} {EOS Trans AGU}\ }\textbf {\bibinfo {volume} {81}},\ \bibinfo
  {pages} {F897} (\bibinfo {year} {2000})}\BibitemShut {NoStop}%
\bibitem [{sup(2021{\natexlab{a}})}]{supplemental-numu}%
  \BibitemOpen
  \href@noop {} {}\bibinfo {howpublished} {See Supplemental Material at [[To be
  inserted by publisher] for the muon neutrino distributions in each quartile
  of hadronic energy fraction} (\bibinfo {year}
  {2021}{\natexlab{a}})\BibitemShut {NoStop}%
\bibitem [{\citenamefont {Acero}\ \emph {et~al.}(2018)\citenamefont {Acero}
  \emph {et~al.}}]{NOvA:2018gge}%
  \BibitemOpen
  \bibfield  {author} {\bibinfo {author} {\bibfnamefont {M.~A.}\ \bibnamefont
  {Acero}} \emph {et~al.} (\bibinfo {collaboration} {NOvA}),\ }\bibfield
  {title} {\bibinfo {title} {{New constraints on oscillation parameters from
  $\nu_e$ appearance and $\nu_\mu$ disappearance in the NOvA experiment}},\
  }\href {https://doi.org/10.1103/PhysRevD.98.032012} {\bibfield  {journal}
  {\bibinfo  {journal} {Phys. Rev. D}\ }\textbf {\bibinfo {volume} {98}},\
  \bibinfo {pages} {032012} (\bibinfo {year} {2018})},\ \Eprint
  {https://arxiv.org/abs/1806.00096} {arXiv:1806.00096 [hep-ex]} \BibitemShut
  {NoStop}%
\bibitem [{\citenamefont {Acero}\ \emph {et~al.}(2020)\citenamefont {Acero}
  \emph {et~al.}}]{Acero:2020eit}%
  \BibitemOpen
  \bibfield  {author} {\bibinfo {author} {\bibfnamefont {M.~A.}\ \bibnamefont
  {Acero}} \emph {et~al.} (\bibinfo {collaboration} {NOvA}),\ }\bibfield
  {title} {\bibinfo {title} {{Adjusting neutrino interaction models and
  evaluating uncertainties using NOvA near detector data}},\ }\href
  {https://doi.org/10.1140/epjc/s10052-020-08577-5} {\bibfield  {journal}
  {\bibinfo  {journal} {Eur. Phys. J. C}\ }\textbf {\bibinfo {volume} {80}},\
  \bibinfo {pages} {1119} (\bibinfo {year} {2020})},\ \Eprint
  {https://arxiv.org/abs/2006.08727} {arXiv:2006.08727 [hep-ex]} \BibitemShut
  {NoStop}%
\bibitem [{\citenamefont {Pinzon~Guerra}\ \emph {et~al.}(2019)\citenamefont
  {Pinzon~Guerra} \emph {et~al.}}]{PinzonGuerra:2018rju}%
  \BibitemOpen
  \bibfield  {author} {\bibinfo {author} {\bibfnamefont {E.~S.}\ \bibnamefont
  {Pinzon~Guerra}} \emph {et~al.},\ }\bibfield  {title} {\bibinfo {title}
  {{Using world charged $\pi^{\pm}-$nucleus scattering data to constrain an
  intranuclear cascade model}},\ }\href
  {https://doi.org/10.1103/PhysRevD.99.052007} {\bibfield  {journal} {\bibinfo
  {journal} {Phys. Rev. D}\ }\textbf {\bibinfo {volume} {99}},\ \bibinfo
  {pages} {052007} (\bibinfo {year} {2019})},\ \Eprint
  {https://arxiv.org/abs/1812.06912} {arXiv:1812.06912 [hep-ex]} \BibitemShut
  {NoStop}%
\bibitem [{\citenamefont {Tanabashi}\ \emph {et~al.}(date)\citenamefont
  {Tanabashi} \emph {et~al.}}]{PhysRevD.98.030001}%
  \BibitemOpen
  \bibfield  {author} {\bibinfo {author} {\bibfnamefont {M.}~\bibnamefont
  {Tanabashi}} \emph {et~al.} (\bibinfo {collaboration} {Particle Data
  Group}),\ }\bibfield  {title} {\bibinfo {title} {Review of particle
  physics},\ }\href {https://doi.org/10.1103/PhysRevD.98.030001} {\bibfield
  {journal} {\bibinfo  {journal} {Phys. Rev. D}\ }\textbf {\bibinfo {volume}
  {98}},\ \bibinfo {pages} {030001} (\bibinfo {year} {2018 and the 2019
  update})}\BibitemShut {NoStop}%
\bibitem [{\citenamefont {Feldman}\ and\ \citenamefont
  {Cousins}(1998)}]{Feldman:1997qc}%
  \BibitemOpen
  \bibfield  {author} {\bibinfo {author} {\bibfnamefont {G.~J.}\ \bibnamefont
  {Feldman}}\ and\ \bibinfo {author} {\bibfnamefont {R.~D.}\ \bibnamefont
  {Cousins}},\ }\bibfield  {title} {\bibinfo {title} {{A Unified approach to
  the classical statistical analysis of small signals}},\ }\href
  {https://doi.org/10.1103/PhysRevD.57.3873} {\bibfield  {journal} {\bibinfo
  {journal} {Phys. Rev. D}\ }\textbf {\bibinfo {volume} {57}},\ \bibinfo
  {pages} {3873} (\bibinfo {year} {1998})},\ \Eprint
  {https://arxiv.org/abs/physics/9711021} {arXiv:physics/9711021
  [physics.data-an]} \BibitemShut {NoStop}%
\bibitem [{\citenamefont {Sousa}\ \emph {et~al.}(2019)\citenamefont {Sousa},
  \citenamefont {Buchanan}, \citenamefont {Calvez}, \citenamefont {Ding},
  \citenamefont {Doyle}, \citenamefont {Himmel}, \citenamefont {Holzman},
  \citenamefont {Kowalkowski}, \citenamefont {Norman},\ and\ \citenamefont
  {Peterka}}]{Sousa:2019nhd}%
  \BibitemOpen
  \bibfield  {author} {\bibinfo {author} {\bibfnamefont {A.}~\bibnamefont
  {Sousa}}, \bibinfo {author} {\bibfnamefont {N.}~\bibnamefont {Buchanan}},
  \bibinfo {author} {\bibfnamefont {S.}~\bibnamefont {Calvez}}, \bibinfo
  {author} {\bibfnamefont {P.}~\bibnamefont {Ding}}, \bibinfo {author}
  {\bibfnamefont {D.}~\bibnamefont {Doyle}}, \bibinfo {author} {\bibfnamefont
  {A.}~\bibnamefont {Himmel}}, \bibinfo {author} {\bibfnamefont
  {B.}~\bibnamefont {Holzman}}, \bibinfo {author} {\bibfnamefont
  {J.}~\bibnamefont {Kowalkowski}}, \bibinfo {author} {\bibfnamefont
  {A.}~\bibnamefont {Norman}},\ and\ \bibinfo {author} {\bibfnamefont
  {T.}~\bibnamefont {Peterka}},\ }\bibfield  {title} {\bibinfo {title}
  {{Implementation of Feldman-Cousins Corrections and Oscillation Calculations
  in the HPC Environment for the NOvA Experiment}},\ }\href
  {https://doi.org/10.1051/epjconf/201921405012} {\bibfield  {journal}
  {\bibinfo  {journal} {EPJ Web Conf.}\ }\textbf {\bibinfo {volume} {214}},\
  \bibinfo {pages} {05012} (\bibinfo {year} {2019})}\BibitemShut {NoStop}%
\bibitem [{sup(2021{\natexlab{b}})}]{supplemental-tab}%
  \BibitemOpen
  \href@noop {} {}\bibinfo {howpublished} {See Supplemental Material at [to be
  inserted by publisher] for the uncertainties on all choices of the mass
  ordering (Normal or Inverted) and upper or lower $\theta_{23}$ octant (UO,
  LO)} (\bibinfo {year} {2021}{\natexlab{b}})\BibitemShut {NoStop}%
\bibitem [{sup(2021{\natexlab{c}})}]{supplemental-contours}%
  \BibitemOpen
  \href@noop {} {}\bibinfo {howpublished} {See Supplemental Material at [to be
  inserted by publisher] for the profiles of these surfaces on the \dmsq and
  \snsq axes as well as the surfaces computed for the inverted hierarchy case}
  (\bibinfo {year} {2021}{\natexlab{c}})\BibitemShut {NoStop}%
\bibitem [{\citenamefont {Abe}\ \emph {et~al.}(2020)\citenamefont {Abe} \emph
  {et~al.}}]{Abe:2019vii}%
  \BibitemOpen
  \bibfield  {author} {\bibinfo {author} {\bibfnamefont {K.}~\bibnamefont
  {Abe}} \emph {et~al.} (\bibinfo {collaboration} {T2K}),\ }\bibfield  {title}
  {\bibinfo {title} {{Constraint on the matter\textendash{}antimatter
  symmetry-violating phase in neutrino oscillations}},\ }\href
  {https://doi.org/10.1038/s41586-020-2177-0} {\bibfield  {journal} {\bibinfo
  {journal} {Nature}\ }\textbf {\bibinfo {volume} {580}},\ \bibinfo {pages}
  {339} (\bibinfo {year} {2020})},\ \bibinfo {note} {[Erratum: Nature 583, E16
  (2020)]},\ \Eprint {https://arxiv.org/abs/1910.03887} {arXiv:1910.03887
  [hep-ex]} \BibitemShut {NoStop}%
\bibitem [{\citenamefont {Abe}\ \emph {et~al.}(2018)\citenamefont {Abe} \emph
  {et~al.}}]{Abe:2017aap}%
  \BibitemOpen
  \bibfield  {author} {\bibinfo {author} {\bibfnamefont {K.}~\bibnamefont
  {Abe}} \emph {et~al.} (\bibinfo {collaboration} {Super-Kamiokande}),\
  }\bibfield  {title} {\bibinfo {title} {{Atmospheric neutrino oscillation
  analysis with external constraints in Super-Kamiokande I-IV}},\ }\href
  {https://doi.org/10.1103/PhysRevD.97.072001} {\bibfield  {journal} {\bibinfo
  {journal} {Phys. Rev. D}\ }\textbf {\bibinfo {volume} {97}},\ \bibinfo
  {pages} {072001} (\bibinfo {year} {2018})},\ \Eprint
  {https://arxiv.org/abs/1710.09126} {arXiv:1710.09126 [hep-ex]} \BibitemShut
  {NoStop}%
\bibitem [{\citenamefont {Adamson}\ \emph {et~al.}(2020)\citenamefont {Adamson}
  \emph {et~al.}}]{MINOS:2020llm}%
  \BibitemOpen
  \bibfield  {author} {\bibinfo {author} {\bibfnamefont {P.}~\bibnamefont
  {Adamson}} \emph {et~al.} (\bibinfo {collaboration} {MINOS+}),\ }\bibfield
  {title} {\bibinfo {title} {{Precision Constraints for Three-Flavor Neutrino
  Oscillations from the Full MINOS+ and MINOS Dataset}},\ }\href
  {https://doi.org/10.1103/PhysRevLett.125.131802} {\bibfield  {journal}
  {\bibinfo  {journal} {Phys. Rev. Lett.}\ }\textbf {\bibinfo {volume} {125}},\
  \bibinfo {pages} {131802} (\bibinfo {year} {2020})},\ \Eprint
  {https://arxiv.org/abs/2006.15208} {arXiv:2006.15208 [hep-ex]} \BibitemShut
  {NoStop}%
\bibitem [{\citenamefont {Aartsen}\ \emph {et~al.}(2018)\citenamefont {Aartsen}
  \emph {et~al.}}]{Aartsen:2017nmd}%
  \BibitemOpen
  \bibfield  {author} {\bibinfo {author} {\bibfnamefont {M.~G.}\ \bibnamefont
  {Aartsen}} \emph {et~al.} (\bibinfo {collaboration} {IceCube}),\ }\bibfield
  {title} {\bibinfo {title} {{Measurement of Atmospheric Neutrino Oscillations
  at 6\textendash{}56 GeV with IceCube DeepCore}},\ }\href
  {https://doi.org/10.1103/PhysRevLett.120.071801} {\bibfield  {journal}
  {\bibinfo  {journal} {Phys. Rev. Lett.}\ }\textbf {\bibinfo {volume} {120}},\
  \bibinfo {pages} {071801} (\bibinfo {year} {2018})},\ \Eprint
  {https://arxiv.org/abs/1707.07081} {arXiv:1707.07081 [hep-ex]} \BibitemShut
  {NoStop}%
\bibitem [{sup(2021{\natexlab{d}})}]{supplemental-dcp}%
  \BibitemOpen
  \href@noop {} {}\bibinfo {howpublished} {See Supplemental Material at [To be
  inserted by publisher] for profiles of these surfaces on the \deltacp axis}
  (\bibinfo {year} {2021}{\natexlab{d}})\BibitemShut {NoStop}%
\bibitem [{\citenamefont {Abe}\ \emph {et~al.}(2021)\citenamefont {Abe} \emph
  {et~al.}}]{T2K:2021xwb}%
  \BibitemOpen
  \bibfield  {author} {\bibinfo {author} {\bibfnamefont {K.}~\bibnamefont
  {Abe}} \emph {et~al.} (\bibinfo {collaboration} {T2K}),\ }\bibfield  {title}
  {\bibinfo {title} {{Improved constraints on neutrino mixing from the T2K
  experiment with $\mathbf{3.13\times10^{21}}$ protons on target}},\ }\href
  {https://doi.org/10.1103/PhysRevD.103.112008} {\bibfield  {journal} {\bibinfo
   {journal} {Phys. Rev. D}\ }\textbf {\bibinfo {volume} {103}},\ \bibinfo
  {pages} {112008} (\bibinfo {year} {2021})},\ \Eprint
  {https://arxiv.org/abs/2101.03779} {arXiv:2101.03779 [hep-ex]} \BibitemShut
  {NoStop}%
\bibitem [{\citenamefont {Kelly}\ \emph {et~al.}(2021)\citenamefont {Kelly},
  \citenamefont {Machado}, \citenamefont {Parke}, \citenamefont
  {Perez~Gonzalez},\ and\ \citenamefont {Zukanovich-Funchal}}]{Kelly:2020fkv}%
  \BibitemOpen
  \bibfield  {author} {\bibinfo {author} {\bibfnamefont {K.~J.}\ \bibnamefont
  {Kelly}}, \bibinfo {author} {\bibfnamefont {P.~A.}\ \bibnamefont {Machado}},
  \bibinfo {author} {\bibfnamefont {S.~J.}\ \bibnamefont {Parke}}, \bibinfo
  {author} {\bibfnamefont {Y.~F.}\ \bibnamefont {Perez~Gonzalez}},\ and\
  \bibinfo {author} {\bibfnamefont {R.}~\bibnamefont {Zukanovich-Funchal}},\
  }\bibfield  {title} {\bibinfo {title} {{Back to (Mass-)Square(d) One: The
  Neutrino Mass Ordering in Light of Recent Data}},\ }\href
  {https://doi.org/10.1103/PhysRevD.103.013004} {\bibfield  {journal} {\bibinfo
   {journal} {Phys. Rev. D}\ }\textbf {\bibinfo {volume} {103}},\ \bibinfo
  {pages} {013004} (\bibinfo {year} {2021})},\ \Eprint
  {https://arxiv.org/abs/2007.08526} {arXiv:2007.08526 [hep-ph]} \BibitemShut
  {NoStop}%
\bibitem [{\citenamefont {Denton}\ \emph {et~al.}(2021)\citenamefont {Denton},
  \citenamefont {Gehrlein},\ and\ \citenamefont {Pestes}}]{Denton:2020uda}%
  \BibitemOpen
  \bibfield  {author} {\bibinfo {author} {\bibfnamefont {P.~B.}\ \bibnamefont
  {Denton}}, \bibinfo {author} {\bibfnamefont {J.}~\bibnamefont {Gehrlein}},\
  and\ \bibinfo {author} {\bibfnamefont {R.}~\bibnamefont {Pestes}},\
  }\bibfield  {title} {\bibinfo {title} {{$CP$ -Violating Neutrino Nonstandard
  Interactions in Long-Baseline-Accelerator Data}},\ }\href
  {https://doi.org/10.1103/PhysRevLett.126.051801} {\bibfield  {journal}
  {\bibinfo  {journal} {Phys. Rev. Lett.}\ }\textbf {\bibinfo {volume} {126}},\
  \bibinfo {pages} {051801} (\bibinfo {year} {2021})},\ \Eprint
  {https://arxiv.org/abs/2008.01110} {arXiv:2008.01110 [hep-ph]} \BibitemShut
  {NoStop}%
\bibitem [{\citenamefont {Chatterjee}\ and\ \citenamefont
  {Palazzo}(2021)}]{Chatterjee:2020kkm}%
  \BibitemOpen
  \bibfield  {author} {\bibinfo {author} {\bibfnamefont {S.~S.}\ \bibnamefont
  {Chatterjee}}\ and\ \bibinfo {author} {\bibfnamefont {A.}~\bibnamefont
  {Palazzo}},\ }\bibfield  {title} {\bibinfo {title} {{Non-standard neutrino
  interactions as a solution to the NO$\nu$A and T2K discrepancy}},\ }\href
  {https://doi.org/10.1103/PhysRevLett.126.051802} {\bibfield  {journal}
  {\bibinfo  {journal} {Phys. Rev. Lett.}\ }\textbf {\bibinfo {volume} {126}},\
  \bibinfo {pages} {051802} (\bibinfo {year} {2021})},\ \Eprint
  {https://arxiv.org/abs/2008.04161} {arXiv:2008.04161 [hep-ph]} \BibitemShut
  {NoStop}%
\end{thebibliography}%

\end{document}